%% file: main.tex
\documentclass[final,5p,times,twocolumn]{elsarticle}

\usepackage{amssymb}
\usepackage{amsmath}

\usepackage{lineno} 
\usepackage{xcolor}
\usepackage{pgfplots}
\usepackage{hyperref}
\usepackage{cleveref}
\usepackage{booktabs}
\usepackage{graphicx}
\usepackage{lipsum}
\usepackage{import}
\usepackage{tikz}
\usetikzlibrary{calc,math, arrows, decorations.pathreplacing,shapes.symbols}
\usetikzlibrary{shapes.geometric}
\usetikzlibrary{arrows.meta}
\usepackage{orcidlink}
\usepackage{pifont}

\journal{Elsevier}

\begin{document}

\include{macros}
\include{tikzset}

\begin{frontmatter}



\title{Data-driven correlations for thermohydraulic roughness properties}


\author[KIT]{Simon Dalpke\corref{cor1} {\orcidlink{0000-0003-0552-9184}}}
\ead{simon.dalpke@kit.edu}
\cortext[cor1]{Corresponding author}
\author[KIT]{Jiasheng Yang {\orcidlink{0000-0003-0091-6855}}}
\author[Aarhus]{Pourya Forooghi \orcidlink{0000-0001-9212-514X}}
\author[KIT]{Bettina Frohnapfel {\orcidlink{0000-0002-0594-7178}}}
\author[KIT]{Alexander Stroh {\orcidlink{0000-0003-0850-9883}}}
\affiliation[KIT]{organization={Institute of Fluid Mechanics, Karlsruhe Institute of Technology (KIT)},
            addressline={Kaiserstraße 12},
            city={Karlsruhe},
            postcode={76131},
            country={Germany}}
\affiliation[Aarhus]{organization={Department of Mechanical \& Production Engineering, Aarhus University},
            addressline={Katrinebjergvej 89},
            city={Aarhus},
            postcode={8200},
            country={Denmark}}

\begin{abstract}
The influence of rough surfaces on fluid flow is characterized by the downward shift in the logarithmic layer of velocity and temperature profiles, namely the velocity roughness function $\Delta U^+$ and the corresponding temperature roughness function $\Delta \Theta^+$. 
Their computation relies on computational simulations, and hence a simple prediction without such simulation is envisioned. 
We present a framework, where a data-driven model is developed using the dataset of Yang et al. 2023 \cite{yang_2023} with $93$ high fidelity direct numerical simulations of a fully-developed turbulent channel flow at $Re_\tau \approx 800$ and $Pr = 0.71$.
The model provides robust predictive capabilities (mean squared error $\text{MSE}_{k} = 0.09$ and $\text{MSE}_\theta = 0.096$), but lacks interpretability. 
Simplistic statistical roughness parameters provide a more understandable route, so the framework is extended with a symbolic regression approach to distill an empirical correlation from the data-driven model. 
The derived expression leads to a predictive correlation for the equivalent sand-grain roughness $k_\text{s} = k_\text{99} (ES_x ( - ES_x + Sk + 2.37) + 0.772)$ with reasonable predictive powers. The predictive capability of the temperature roughness function is subject to limitations due to the missing Prandtl-number variation in the dataset. 
Nevertheless, the interpretable correlation and the neural network as well as the original dataset can be used to explore the roughness functions. 
The functional form of the derived correlations, along with visual analysis of these surfaces, suggests a strong relationship with roughness wavelengths, further linking them to explanations based on sheltered and windward regions.
\end{abstract}


\end{frontmatter}

\section{Introduction}
\label{sec:Introduction}
Rough surfaces are omnipresent in real-world applications, e.g., rough walls of heat exchanger due to additive manufacturing \cite{garg_2024} or the effect of debris and corrosion on gas turbines \cite{bons_2010}. This in turn influences the fluid flow with severe impact on the heat exchange capabilities and friction force. 
Roughness is an inherently multiscale phenomenon, encompassing a broad range of length scales that define a rough surface.
This multiscale nature of rough surfaces can be linked to various origins, ranging from the particular deta
Eventually, this multiscale phenomenon influences the fluid flow and the resultant time- and space-averaged velocity profile above these rough surfaces is of interest, so the rough and smooth velocity profile under the same flow conditions can be compared. The roughness influence is represented by the shift of the logarithmic intercept $\Delta U^+$ in the logarithmic region \cite{hama_1954, clauser_1954}. Using the work by \citet{nikuradse_1933}, this velocity roughness function $\Delta U^+$ can be expressed in terms of the equivalent sand-grain roughness by comparing the investigated surface with a roughness consisting of spherical elements with diameter $k_\text{s}$ at the same friction factor in the fully rough regime -- a condition, in which $\Delta U^+$ can be determined by a single length scale. 
Hence, the velocity roughness functions read
\begin{equation}
    \Delta U^+ = \frac{1}{\kappa} \ln \lb k_\text{s}^+ \rb - A + B
\end{equation}
where $B\approx 5.2$ is the smooth wall logarithmic intercept and $A$ the rough wall equivalent.
The fundamental question is, how this single length scale can be determined solely based on the roughness height distribution or some chosen roughness parameters and hence without costly simulations necessary for comparison. It should be clearly noted, that the equivalent sand-grain roughness is a hydrodynamic property rather than a geometric one. 
Hence, the envisioned \textit{a priori} determination should link the roughness description to the physics of flow disturbed by the rough wall. 

Two approaches towards a relation between the equivalent sand-grain roughness and the rough surface are omnipresent in literature. Over the last decades, multiple empirical correlations have been established using scalar roughness parameters (an extensive collection can be found in Ref. \cite{abdelaziz_2024}), whereas in recent years, data-driven approaches were developed more frequently. 
The first approach builds on simple and understandable parameters, such as the mean height $k_\text{avg}$ or the skewness $Sk$ computed from the height profile $h(x_1,x_3)$. 
These correlations are often based on experimental campaigns (e.g., \cite{flack_2016,flack_2020}) or extensive direct numerical investigations (e.g., \cite{forooghi_2017,kuwata_2019}) and varying sources for the roughness, e.g., \citet{flack_2016} uses different grit-blasted surfaces or \citet{bons_2002} investigates roughness on turbine blades. Examples of these mentioned correlations as well as the simplistic correlation by \citet{chan_2015} are presented in \cref{tab:example-correlation}, covering numerical simulations and experimental campaigns as well as different roughness types. Shared characteristics of these correlations are the understandable nature and the simple geometrical parameters used. A classical approach to develop a roughness correlation builds on systematic variations of roughness parameters, e.g., different skewness values by \citet{flack_2020} or the usage of geometrical simple roughness elements to achieve certain values for roughness parameters \cite{forooghi_2017}. New roughness parameters were repeatedly proposed with the aim of finding meaningful correlations, and a summary of these topological parameters is given by \citet{chung_2021}. A clear picture on the most informative parameters remains unknown, but a measure of height, frontal area and slope are favorable \cite{flack_2022a}.
\citet{chung_2021} concluded that, a universal correlation remains difficult to establish, especially as most correlations are designed for a limited roughness parameter space and further improvement will be possible by carefully investigate certain topological features. 

On the other hand, data-driven approaches became more valuable and more frequently used in recent years;  \citet{jouybari_2021}, \citet{yang_2023}, and \citet{sanhueza_2023a} can be named as examples. These examples make use of the elaborated approximation behavior of neural networks, which leads to impressive predictive capabilities. In return, this requires a high dimensional input. 
For example, \citet{jouybari_2021} requires 17 geometrical roughness parameters to predict the normalized equivalent sand-grain roughness using a simple multi-layer perceptron (MLP). Similar output with different input vector is employed by \citet{yang_2023} building on the probability density function as well as the power spectrum to incorporate the multi-scale nature of rough surfaces. A different approach is conducted by \citet{sanhueza_2023a} utilizing a convolutional neural network to analyze roughness height maps and predict local drag and heat transfer behavior. All in all, these data-driven approaches require detailed information on the rough surfaces, complicating their use in engineering applications. 
Hence, the question arises, how the sophisticated neural networks can be transferred into simpler and easy-to-apply correlations, thereby sacrificing predictive capabilities but enhancing the applicability in engineering applications. 
\begin{table*}
  \centering
  \ifpreprint
    \hspace*{-0.9cm}%
  \fi
  \begin{tabular}{c c}
    \toprule
    Reference & Correlation \\
    \midrule 
    \vspace{5pt}
    \citet{chan_2015} & $k_{\text{s}} = 7.3 ES^{0.45} k_{\text{a}}$ \\
    \vspace{5pt}
    \citet{flack_2016} & $k_{\text{s}} = 3.47 k_{\text{rms}} \left(Sk + 2\right)^{0.405} $ \\
    \vspace{5pt}
    \citet{forooghi_2017} & $k_{\text{s}} = \bar{k}_{\text{t}} \left(1.07 - 1.07 e^{- 3.5 ES}\right) \left(0.67 Sk^{2} + 0.93 Sk + 1.3\right)$ \\
    \vspace{5pt}
    \citet{kuwata_2019} & $k_{\text{s}} = 4 k_{\text{rms}} (0.17 Sk + 1)^4$ \\
    \citet{flack_2020} & $k_{\text{s}} = \begin{cases} 2.48 k_{\text{rms}} \left(Sk + 1\right)^{2.24} & \text{for}\: Sk > 0 \\\frac{2.73 k_{\text{rms}}}{\left(Sk + 2\right)^{0.45}} & \text{for}\: Sk < 0 \\2.11 k_{\text{rms}} & \text{otherwise} \end{cases}$ \\
    \bottomrule
  \end{tabular}
  \caption{Selection of correlations for the equivalent sand-grain roughness from various literature sources.}
  \label{tab:example-correlation}
\end{table*}

For temperature (passive scalar) considerations, e.g., for analyzing the heat exchange influence, the picture further complicates. 
Similarly to the velocity profile, the spatial and temporal averaged temperature profile shifts in the logarithmic region, but the presence of the Prandtl-number $Pr$ as a material parameter adds further complexity.
As summarized by \citet{chung_2021} or \citet{kadivar_2024a} depending on the wall-normal scaling, multiple relations for the rough wall temperature profile can be derived. As one example, following the concept of the equivalent sand-grain roughness, the temperature roughness function follows
\begin{equation}
  \Delta \Theta^+ = \frac{1}{\kappa_\theta} \ln \lb k_\text{s}^+ \rb + A_\theta(Pr) - g(k_\text{s}^+, Pr),
  \label{eq:delta-theta-g}
\end{equation}
using the roughness-dependent $g$-function \cite{dipprey_1963}, the von Kármán constant for heat transfer $\kappa_\theta$, and $A_\theta$ being the smooth wall logarithmic intercept \cite{kader_1981}. As collected by \citet{kadivar_2024a} multiple correlation usually relating the $g$-function or $\Delta \Theta^+$ to the equivalent sand-grain roughness $k_\text{s}^+$ and the Prandtl-number $Pr$ exist, but a universal equation remains unidentified. Additionally, in contrast to the fully rough regime for the velocity roughness function, no clear asymptotic behavior is discovered for the thermal roughness function yet. 
Various trends have been reported so far, ranging from constant increase over leveling-off behavior to a decrease of $\Delta \Theta^+$ for increasing $k_\text{s}^+$ \cite{hantsis_2024,kadivar_2024a}. \citet{hantsis_2024} further conclude, based on the reviewed publications, that the flattening of $\Delta \Theta^+$ for increased $k_\text{s}^+$ suggest an asymptotic value of the temperature roughness function, influenced by Prandtl-number and roughness properties -- so a clear picture is yet to be developed.
Similar to the velocity roughness function, neural networks can be used to predict the temperature roughness function as well, and understanding the behavior of the network through a symbolic expression might help in developing a picture for the roughness influence on temperature.

In general, neural networks are an effective tool for function approximation, but lack in interpretability due to their convoluted structure. In contrast, a simplistic mathematical expression with well-understood variables is able to describe a given phenomenon in a human-interpretable manner. The relation between the neural network and an interpretable correlation can be established using \textit{symbolic regression} -- a model regression analysis for development of a mathematical expression, which describes an unknown relation inside the data \cite{angelis_2023}. Early studies like \cite{gerwin_1974, langley_1981} started using data to develop mathematical expressions, but the work of \citet{koza_1994} on genetic programming broadened the application spectrum, as summarized in \cite{cranmer_2023}. 
Recent successful applications of symbolic regression in fluid mechanics range from deriving expression for Stokes flow around multiple particles \cite{reuter_2022} to turbulence modelling \cite{schmelzer_2020}.\\

In this contribution, we aim to bridge the gap between the data-driven predictive methods (neural networks) and understandable roughness correlations by applying a symbolic regression approach to transfer the first into a correlation.
Therefore, we utilize an extensive database (described in section \ref{sec:database}) to train and test neural networks for the prediction of the temperature and velocity roughness function (section \ref{sec:neural-networks}). These networks are then deployed on a set of $4200$ artificial rough surfaces, which forms the base for a symbolic regression approach (section \ref{sec:symbolic-regression}), to obtain interpretable mathematical expressions. The developed predictive methods, namely the neural networks and the evolved correlations, together with the results from direct numerical simulations are investigated and explored in section \ref{sec:results}.

\section{Database}
\label{sec:database}
The starting point for development of thermohydraulic correlations is a comprehensive database. Here, we use the database by \citet{yang_2023} with $4200$ artificial generated rough surfaces following a validated generation algorithm by~\citet{perez-rafols_2019}, which uses the probability density function (PDF) and the power spectrum (PS) of the roughness height values $h(x_1,x_3)$, where subscript  $1$ and $3$ represents wall-parallel streamwise and spanwise coordinates, respectively, to generate an artificial surface mimicking realistic conditions. These two measures are known to contain the most information of the rough surface \cite{persson_2023} and provide a nearly unique relation to the velocity roughness function \cite{yang_2022a}. As summarized by \citet{persson_2023}, the PS and PDF are further related to a set of scalar statistical roughness parameters, e.g., the root-mean-square height $k_\text{rms}$. The $4200$ artificial rough surfaces span a maximum height range $0.06 < k_\text{t}/\delta < 0.18$, where $\delta$ is the half channel height of the fluid flow simulation and $k_\text{t}$ is the peak-to-through height. As the roughness height map is normalized using the channel half height $\delta$, the dimension-related statistical variables (e.g., $k_\text{rms}$ or $k_\text{avg}$) are normalized by the channel half height as well, which will not be stated explicitly throughout the study for brevity.

In the database, a subset of the $4200$ rough surfaces were evaluated at $Re_{\tau} \approx 800$ in a periodic channel setup using direct numerical simulation carried out with the pseudo-spectral code SIMSON~\cite{chevalier_2007} to numerically evaluate the Navier-Stokes-Equations at a constant pressure gradient condition augmented with a passive scalar for the temperature field ($Pr=0.71$):
\begin{align}
  \pd{u_i}{x_i} &= 0, \\
  \pd{u_i}{t} + u_j \pd{u_i}{x_j} &= - \pd{p}{x_i} + \frac{1}{Re} \pdm{2}{u_i}{x_j}{x_j} - \Pi \fe_1 + \ff_{IBM},\\
  \pd{\theta}{t} + u_i \pd{\theta}{x_i} &= \alpha \pdm{2}{\theta}{x_i}{x_i} - u_1 \td{\theta_w}{x_1} + \ff_{IBM}.
\end{align}
The additional terms $\Pi$ and $u_1 \d \theta_w / \d x_1$ enforce a constant pressure gradient and a statistically uniform wall heat flux at the wall respectively. The roughness is symmetrically applied at the upper and lower wall so that the deepest trough is located at $y=0$ and $y=2\delta$ with the half channel height $\delta$ and introduced in the simulation domain by an immersed boundary method~\cite{goldstein_1993}. 
The high computational cost to generate a detailed database were reduced by applying the minimal channel approach \cite{jimenez_1991} transferred to rough surfaces by \cite{chung_2015} and \cite{macdonald_2016}. 
With this approach, the solution remains valid in the near-wall region and is sufficient to compute the roughness function, while a departure from the turbulent solution is expected above the critical height $y_{\text{crit}}^+ = 0.4 L_z^+$, which is linked to the insufficient capturing of the large scales of turbulent flow. 
An active learning loop was implemented to optimize computational effort and enhance efficiency during the database creation by \citet{yang_2023}. Here, an ensemble of neural networks was trained using an initial subset of simulations and based on the evaluated uncertainty of the neural network prediction selected rough surfaces were simulated \cite{yang_2023}. 
The simulation workflow is validated in~\citet{yang_2023d}.
Normalization of the velocity and temperature profile are performed using the wall friction velocity $u_{\tau} = \sqrt{\tau_\text{w} / \rho}$ and the friction temperature $\theta_{\tau} = \langle q_\text{w} \rangle_\text{w} / (\rho c_\text{p} u_{\tau})$. Here $\langle \cdot \rangle_w$ describes the average over the wetted area $A_\text{w}$ normalized by the plane area $A$ of the channel. Due to the used boundary conditions, the wall heat flux $\langle q_\text{w} \rangle_\text{w}$ can be determined through energy conservation consideration and similarly $\tau_\text{w}$ follows from force equilibrium.

The database furthermore contains smooth wall reference simulation with similar minimal channel dimensions. These simulations are used to compare smooth and rough wall simulations in the logarithmic region to derive the velocity and temperature roughness function. The roughness functions are evaluated by averaging the difference between rough and smooth profile in the region $160 < (y-d)^+ < 400$.
Jackson height \cite{jackson_1981} is used as the virtual origin during evaluation of both roughness functions ($d=d_{\theta}$), whereas the prescribed constant pressure gradient is computed based on the average channel height a priori to the simulation ($\Pi = \tau_\text{w} (\delta - k_{\text{avg}})$). Here $k_{\text{avg}}$ is the average height of the roughness profile, measured from its deepest trough, also known as meltdown height. 
Further information on the database is presented in \cite{yang_2023} as well as in \ref{app:database}.

The estimation of the equivalent sand-grain roughness from $\Delta U^+$ is performed using
\begin{equation}
  k_{\text{s}}^+ = \exp \lb \kappa \lb \Delta U^+  + A - B \rb \rb .
\end{equation}
In the fully rough regime ($k_\text{s}^+\geq70$ \cite{flack_2010}) the rough wall log intercept reaches a finite value independent of roughness Reynolds number $A \rightarrow 8.48$. \citet{yang_2023} observed, that including surfaces with $50 \leq k_\text{s}^+ \leq 70$ improved the model performance. 
Hence, only simulations with $\Delta U^+ < 6$ (relating to $k_\text{s}^+ < 45$) are discarded from the dataset. This lead to a dataset of $93$ simulations, where $79$ are in the fully rough regime ($k_\text{s}^+ \geq 70$). Eleven simulations are in the range $50 \leq k_\text{s}^+ \leq 70$, and two additional data points are added to the dataset by only excluding $\Delta U^+ < 6$ simulations.
A dimensionless version of the equivalent sand-grain roughness is generated analog to \citet{yang_2023} via $k_\text{r} = k_\text{s}/k_{99}$, where $k_{99}$ is the $99\%$ confidence interval of the probability density function of the height map. For the temperature roughness function, such a normalization is not easily available, so we consider only the $\Delta \Theta^+$ as a predictive quantity.

The dataset is then split into three sets for different tasks during the data-driven procedure. The training set $\cD$ contains $60$ simulations and is used during training of neural networks. During the hyperparameter search, $16$ data points are used as validation dataset $\cV$. Finally, the test set $\cT$ is used to assess the generalization capabilities and contains $16$ DNS results. The split of the dataset is performed using a pseudo-random assignment (using \texttt{scikit-learn} \cite{pedregosa_2011}).

Additionally, two external datasets $\cE$ are ($\cE_k$for $k_\text{r}$ containing $20$ samples and $\cE_\Theta$ for $\Delta \Theta^+$ made up by $8$ samples) curated with rough surfaces from an online roughness database (\url{www.roughnessdatabase.org}) to evaluate the generalization capabilities. These surfaces range from surface scans to artificial rough surfaces created by randomly placing roughness elements of a certain geometrical type.
Please note, that the rough surfaces in the external dataset were not generated using the algorithm described by \citet{perez-rafols_2019} but consist of a great variety of surfaces, ranging from surface scans to randomly placed ellipsoids. A short review of the topological parameters of the datasets is given in \ref{app:database}. For the comparison we use the equivalent sand-grain roughness reported in this database, but for the temperature roughness function we had to perform additional simulations with selected surfaces according to the procedure described above. To enlarge the external dataset for the temperature roughness function, we add the simulation results by \citet{yang_2023d} of realistic ice accretion.

\section{Neural Networks for Roughness Function Prediction}
\label{sec:neural-networks}
As a reference for the neural network, a simplistic baseline model is developed. It represents the most simple model and if no improvements of the more complex neural network is achieved, the baseline model should be preferred. For that, a linear regression model is implemented using \texttt{scikit-learn} \cite{pedregosa_2011} and trained on the validation and training dataset. Evaluated on the test dataset, the linear regression model archives a mean squared error (MSE, average squared difference between predicted and actual values)  $\text{MSE} = 0.614$ and $0.744$ for the prediction of $k_{\text{r}}$ and $\Delta \Theta^+$ respectively. This result indicates a fairly limited predictive ability for the baseline model, further highlighted by the mean absolute percentage error (MAPE, average percentage deviation of predictions from actual values)  $\text{MAPE} = 37.6\%$ for the prediction of $k_{\text{r}}$. \\

The input of both neural networks predicting the normalized equivalent sand-grain roughness and the temperature roughness function follows \citet{yang_2023} and consists of the discretized power spectrum (PS) and the discretized probability density function (PDF) as well as three scalar roughness parameters ($k_\text{t}/k_{99}$, $\lambda_0/k_{99}$ and $\lambda_1/k_{99}$). PS and PDF are sampled at $30$ equidistantly distributed points inside the ranges:
\begin{align}
  \text{PDF:} \quad & 0 < k < k_\text{t}\\
  \text{PS:} \quad & \lambda_0 < \lambda < \lambda_1
\end{align}
It showed furthermore beneficial to normalize the input data by rescaling the input features to the range $[0,1]$, preserving relative relationships while mitigating the influence of various feature magnitudes. The final roughness feature vector $x \in \ffR^{63}$ serves as input layer to the neural network. 
The architecture of the networks starts with a fully connected hidden layer containing $n_\text{n}$ neurons. This is followed by $n_\text{h}$ layers consisting of half of the neurons of the preceding layer. The dropout rate, activation function, batch size, kernel initializer as well as the scaling factor of $L^2$-kernel regularization penalty are introduced as hyperparameters in the architecture. Finally, a single output neuron with linear activation function is used to output $k_\text{r}$ or $\Delta \Theta^+$. It should be mentioned, as no Prandtl-number and Reynolds-number related quantity are included in the network, the generalization capability for $\Delta \Theta^+$ is limited, especially as literature resources (e.g., the simulation campaign by \citet{zhong_2023} or the literature review by \citet{hantsis_2024}) heavily suggest such dependencies. 
During the training phase, the loss function is the mean-squared error MSE, and the Adam-Optimizer with a specified learning rate is used to govern the backpropagation algorithm.
To obtain values for the above specified hyperparameters, a hyperparameter optimization using a random search algorithm is implemented using \texttt{KerasTuner} \cite{omalley_2019}. 
To improve the predictive capabilities, values for the hyperparameters are randomly sampled from a specified search space. The resulting architecture is trained using the training set $\cD$ three times to account for random effects during initialization and evaluated on the validation set $\cV$. The obtained architecture can then be statistically evaluated and deployed. 

Interestingly, the hyperparameter search for the normalized equivalent sand-grain roughness and the temperature roughness function lead to identical values for both networks. Networks containing $5$ hidden layers and $128$ start neurons, the sigmoid linear unit as activation function and $10\%$ dropout rate, were found to be the most effective. The reason, why the same architecture is evolved by the hyperparameter search remains to be elucidated, but the authors hypothesize that it may be associated with the similarities in the dataset or the nature of the output values.
Furthermore, the hyperparameter optimization can unveil trends, e.g., a high batch size and high learning rate showed to be unfavorable for the prediction of $k_\text{r}$. 

The architecture of the final hyperparameter search is used to train a deployment version of both neural networks. This version will be used in the following symbolic regression step, if the predictive capabilities exceed the baseline model. 
\begin{figure}
  \centering
  \input{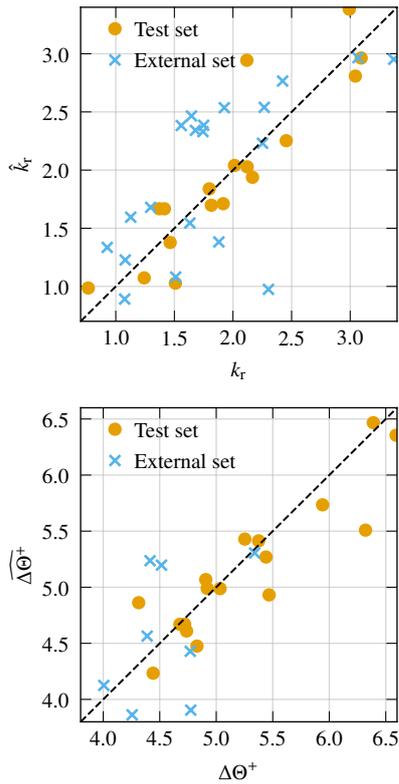}
  \caption{Evaluation of the neural networks for prediction of $k_r$ and $\Delta \Theta^+$ using the test dataset, as well as external simulation results.}
  \label{fig:ResultNN}
\end{figure}
The networks are evaluated using the test dataset $\cT$ and the external dataset, and their results are visualized in \cref{fig:ResultNN}. In \cref{fig:ResultNN}, the true values $(\cdot)$ are compared with the predicted values $(\hat{\cdot})$ from the neural network. It is clearly visible, that the test set for $\Delta \Theta^+$ and $k_\text{r}$ is predicted well, whereas the predictive capabilities for the external set are diminished. This trend is reflected in the scalar metrics as well. In this case, the network optimized through the conducted hyperparameter search predicting $k_\text{r}$ yields in a MAPE of $13.4\%$ on $\cT$, whereas a MAPE of $27.5\%$ is achieved on the external dataset. Similarly, the temperature roughness function prediction follows with a MAPE of $4.3\%$ if evaluated on $\cT$ and $9.5\%$ if evaluated on the external reference data. This observation clearly shows the generalization capabilities of both networks: The external data poses a challenge for the neural networks, e.g., visible in the scalar metrics and the data points with large deviations in \cref{fig:ResultNN}. Nevertheless, the predictions are clearly improved compared to the baseline model. This conclusion coincides with the observation by \citet{yang_2023} on the same dataset, where only the normalized equivalent sand-grain roughness is investigated. They further conclude that the prediction quality is diminished due to the fact, that some rough surfaces of the external dataset might represent unseen roughness types. 
The mean-squared error (MSE) values of $0.09$ and $0.096$ evaluated on the test set $\cT$ for $k_\text{r}$ and $\Delta \Theta^+$ respectively, substantiate the predictive capability of the networks. The slightly higher MSE value for the temperature roughness function might be related to the structure of the training dataset -- the dataset is created using an active learning loop, which is governed through the normalized equivalent sand-grain roughness \cite{yang_2023}. Hence, the parameter space for the temperature roughness function might be underexplored. Furthermore, it is important to acknowledge the limitations of the dataset and architecture used for the temperature roughness function. These constraints restrict the generality of the predictive framework for \(\Delta \Theta^+\), as the training data and model architecture are specific to \(Pr = 0.71\) and \(Re_{\tau} \approx 800\). 

\section{Symbolic Regression}
\label{sec:symbolic-regression}

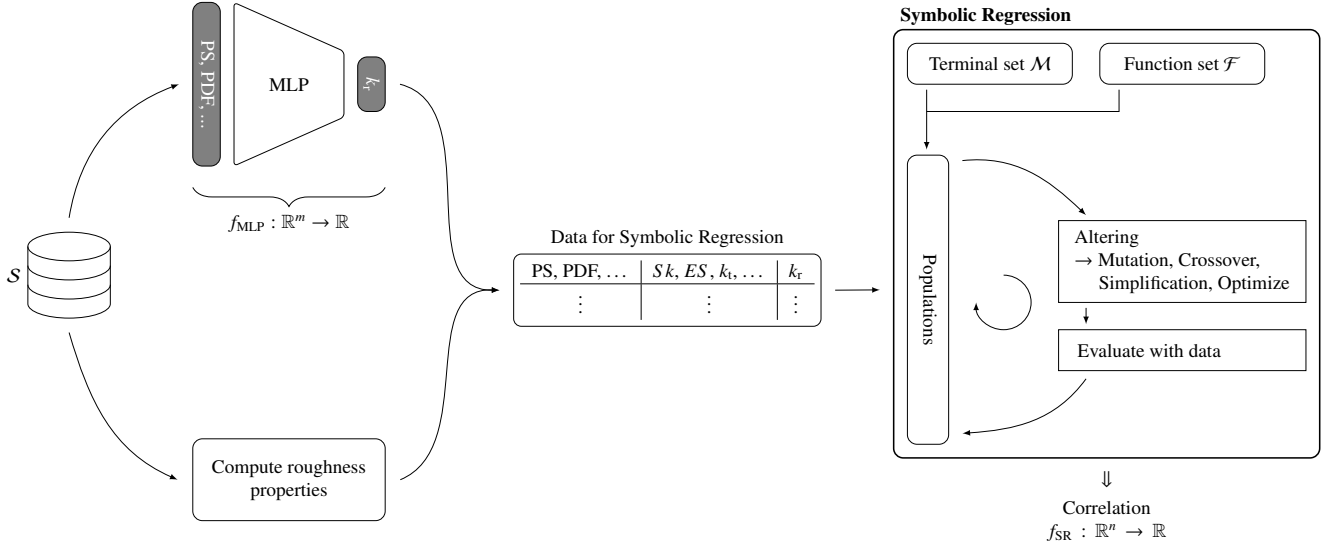
\begin{figure*}
  \centering
  \ifpreprint
    \footnotesize
    \hspace*{-2cm}%
    \resizebox{1.25\textwidth}{!}{\input{figures/symbolic_regression.tex}}
  \else
    \resizebox{0.95\textwidth}{!}{\input{figures/symbolic_regression.tex}}
  \fi
  \caption{Procedure to transfer the neural network into a roughness correlation for the equivalent sand-grain roughness using simplistic roughness parameters.}
  \label{fig:sketch-sr}
\end{figure*}

The developed machine learning models are well suited for predictive tasks, but lack in physical interpretability and simple engineering application. Therefore, it is aimed to transfer the models in a simplistic correlation. The target of these correlation lays in predicting the thermohydraulic properties, e.g., the normalized equivalent sand-grain height, based on measured topological features \cite{chung_2021}. A symbolic regression algorithm is deployed, building on ideas from biological evolution and genetic programming (see \citet{kruse_2015} for a detailed introduction).

Fig. \ref{fig:sketch-sr} visualizes this procedure combining the neural network prediction, the computation of statistical roughness parameters and the symbolic regression approach. Starting with the database $\cS$ containing $4200$ artificial rough surfaces, each roughness is processed through the developed neural network to obtain the normalized equivalent sand-grain roughness or the temperature roughness function. The network itself is a continuous function $f_{\text{MLP}}: \ffR^m \rightarrow \ffR$ relating the input to the scalar output. Additionally, each surface is evaluated to receive the specific roughness parameters following the definitions presented in \citet{chung_2021} and shown in \ref{app:database}. The prediction of $k_\text{r}$ and $\Delta \Theta^+$ together with the scalar roughness parameters form the input to the symbolic regression approach and can be understood as a discretized version of $f_{\text{MLP}}$.

The symbolic regression approach follows an evolutionary algorithm, where a population of mathematical expressions is evolved through genetic operators to fit a dataset, similar to biological evolution. To encode the mathematical expression, genetic programming is applied, which is constrained by a set of rules (grammar) adapted to the given problem \cite{kruse_2015}. The terminal set $\cM$, including input variables and numeric constants, and the function set $\cF$ with primitive functions form this grammar \cite{kruse_2015}. Both sets are necessary to form mathematical expressions as genetic programs. As \citet{kruse_2015} states, the unknown function in the dataset can only be discovered if the sets are sufficiently expressive, or, in other words, a sufficient symbolic expression cannot be assembled if the necessary building blocks are missing.

The \texttt{python} package \texttt{PySR} \cite{cranmer_2023} is used throughout this study. \texttt{PySR} provides a multi-population evolutionary algorithm adapted to mathematical expressions. The populations of mathematical expressions build from terminal and function set are then evolved in an adapted evolution loop. This includes the application of genetic operators (mutation or crossover) as well as simplification and optimization of the equations. A detailed description and further adaptions of \texttt{PySR} are described in \citet{cranmer_2023}. The fitness (loss function) of each evolved equation during evolution is evaluated using the data provided to the symbolic regression algorithm. The evolutionary algorithm stops at a given termination criterion and finally provides a functional relationship between selected scalar parameters and the normalized equivalent sand-grain roughness. 
Hence, the multi-objective optimization problem describing this workflow can be formulated as: 
\begin{quotation}
  \noindent
  Find a function $f_{\text{SR}}: \ffR^n \rightarrow \ffR$ which approximates the discretized version of the function $f_{\text{MLP}}: \ffR^m \rightarrow \ffR$ by minimizing the fitness function and model complexity.
\end{quotation}
The symbolic distillation for the temperature roughness function follows the same procedure.

To apply the algorithm described above and visualized in \cref{fig:sketch-sr}, the terminal and function set have to be defined.
As a large set of terminal features poses a challenge for the symbolic regression process and the observed dependency on the terminal set, a small and large terminal set is used throughout the study. This approach hence combines exploring a wide range of possible equations and exploitation, focusing on refinement and improvement of correlations. 

\begin{table}
    \centering
    \begin{tabular}{c|c c c c}
        \toprule
        Parameter & $\cM_{\text{velo}}^\text{small}$  & $\cM_{\text{velo}}^\text{large}$  & $\cM_{\text{temp}}^\text{small}$ & $\cM_{\text{temp}}^\text{large}$ \\
        \midrule
        $k_\text{avg}/k_\text{rms}$ & \ding{54} & \ding{54} & \ding{54}  & \ding{54}  \\
        $k_\text{t}/k_\text{rms}$ &  & \ding{54}  &  & \ding{54}  \\
        $\bar{k}_\text{t}/k_\text{rms}$ & \ding{54}  & \ding{54}  & \ding{54}  & \ding{54}  \\
        $Sk$ & \ding{54}  & \ding{54}  & \ding{54}  & \ding{54}  \\
        $ES_x$ & \ding{54}  & \ding{54}  & \ding{54}  & \ding{54}  \\
        $Ku$ &  & \ding{54}  &  & \ding{54}  \\
        $\xi_\text{rms}$ &  & \ding{54}  &  & \ding{54}  \\
        $V_\text{mat}$ &  & \ding{54}  & & \ding{54}  \\
        $AC_\text{len}$ &  & \ding{54}  &  & \ding{54}  \\
        $AC_x$ &  & \ding{54}  &  & \ding{54}  \\
        $Re_\tau$ &  &  & \ding{54}  & \ding{54}  \\
        $Pr$ &  &  & \ding{54}  & \ding{54}  \\
        $\hat{k}_\text{s}$ &  &  & \ding{54}  & \ding{54}  \\
        \bottomrule
    \end{tabular}
    \caption{Overview on the terminal symbols used in for the small and large set to determine a correlation for $k_\text{s}$ and $\Delta \Theta^+$}
    \label{tab:terminal_symbols}
\end{table}

For a sufficient prediction of the equivalent sand-grain roughness, a measure for size, coverage and frontal area is necessary \cite{flack_2022a}. Skewness $Sk$, effective slope $ES_x$ as well as $k_{\text{a}}/k_{\text{rms}}$ (average height) or $\bar{k}_{\text{t}}/k_{\text{rms}}$ (mean subdivided peak-to-valley height) provide this information for artificial rough surfaces and are the content of the small terminal set $\cM_{\text{velo}}$. The root-mean-square height is chosen for normalization as it is widely used and simple to determine, which is deemed reasonable for engineering applications.
In the terminal set $\cM_{\text{temp}}$ the friction Reynolds number $Re_{\tau}$, Prandtl number $Pr$ and equivalent sand-grain roughness $\hat{k}_\text{s}$ (predicted by the neural network) were added due to the occurrence in literature correlations, as summarized in Ref. \cite{kadivar_2024a}.
Additional parameters (e.g., autocorrelation length or the root-mean-squared slope) are used in the large terminal set to provide the exploration capabilities. A summary of the roughness parameters in the terminal set is given in \cref{tab:terminal_symbols} and their computation can be found in \ref{app:database}.

\begin{table}
    \centering
    \begin{tabular}{c c}
      \toprule
      Function & Complexity \\
      \midrule
      $+$ & $1$ \\
      $-$ & $1$ \\
      $*$ & $1$ \\
      $/$ & $1$ \\
      \footnotesize{$(\cdot)^{(\cdot)}$}  & $2$  \\
      $\ln$ & $3$ \\
      $\log$ & $3$ \\
      $\exp$ & $3$ \\
      $\abs{\cdot}$ & $3$ \\
      $\sqrt{\cdot}$ & $3$ \\ 
      $(\cdot)^2$ & $3$ \\ 
      $\max$ & $4$ \\
      $\min$ & $4$ \\
      \bottomrule
    \end{tabular}
    \caption{Complexity attributed to the functions in $\cF$.}
    \label{tab:Complexity_Functions}
\end{table}

The second part of setting up the symbolic regression algorithm includes the definition of the function symbols as well as their respective complexity penalties. These are summarized in \cref{tab:Complexity_Functions} and included, in increasing complexity order, basic arithmetic operations, advanced mathematical operations (e.g., $\ln(\cdot)$) as well as the maximum and minimum operator for piecewise correlations. This selection is based on literature observation, e.g., a piecewise function was formulated by \citet{flack_2022} or no trigonometric functions are employed in common literature correlations (see summary of \citet{abdelaziz_2024}). 
Variables of the terminal set, including normalized variables, and constants used in the procedure are of complexity one. The vast majority of the additional hyperparameter of the symbolic regression approach in \texttt{PySR} remained unaffected, but it showed beneficial to increase the number of populations and the population size, as well as extend the number of iterations. Furthermore, the maximal complexity is increased, due to appearance of complex correlations in the literature. The $L^2$ norm is used as a fitness function to evaluate the individual correlations in the population.

To account for random effects in the described algorithm, e.g., the initialization of the population, five executions of the regression algorithm were performed per terminal set. The symbolic regression approach provides a set of equations, consisting of the best equation \cite{cranmer_2023} and promising candidates. It is noteworthy that across multiple executions, identical or similar equations are generated (e.g., the best equations for the normalized equivalent sand-grain roughness in the small terminal coincide for all executions). Important observations in the set of possible equations from the symbolic regression include the occurrence of similar building blocks, e.g., $(Sk + a)$ which are present in literature roughness correlations as well \cite{flack_2016, kuwata_2019}. Furthermore, the large terminal set consistently resulted in higher complexity values of the equation, which only lead to small improvements in the mean squared error.

\begin{figure*}
  \centering
  \ifpreprint
    \hspace*{-2.2cm}%
  \fi
  \input{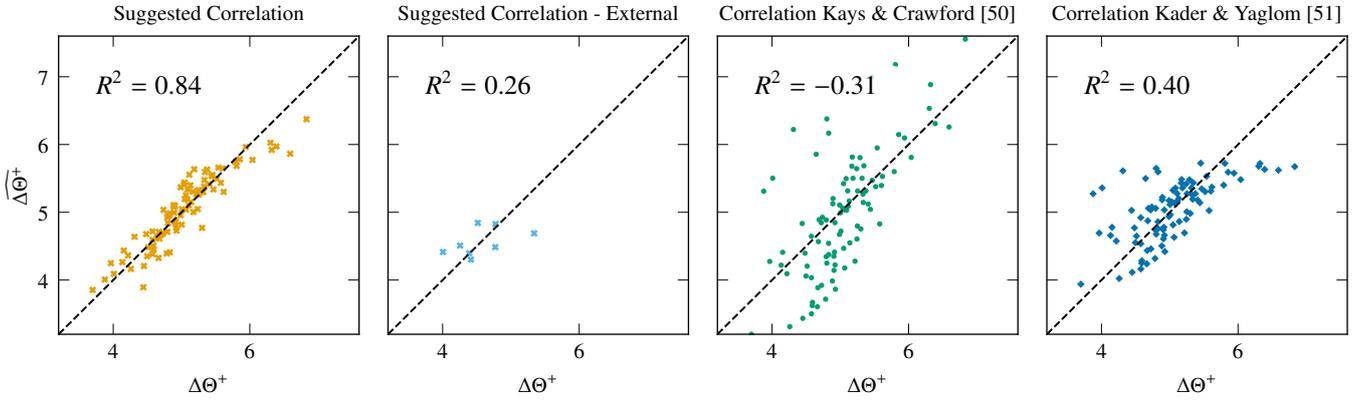}
  \caption{Evaluation of the true $(\cdot)$ and predicted $\hat{(\cdot)}$ values of the temperature roughness function using the dataset of \citet{yang_2023}. The suggested correlation was tested on the external dataset $\cE_\Theta$ as well.}
  \label{fig:compare_temperature}
\end{figure*}

Based on a pairwise comparison of the complexity values and the fitness values as well as known behaviors (e.g., positive skewed surfaces exhibits higher values of $k_\text{s}$ \cite{flack_2020, kuwata_2023}) finally results in a data-driven discovery of a roughness correlation:
\begin{equation}
  k_{\text{r}} = \frac{k_{\text{s}}}{k_{99}} =  ES_x (-ES_x +Sk+2.37)+0.772 , \quad \text{and}
  \label{eq:result_sr_k}
\end{equation}
\begin{equation}
  \Delta \Theta ^+ = 6.022 \lb k_{\text{s}} \lb -0.182 Sk + \frac{\bar{k}_{\text{t}}}{k_{\text{rms}}} \rb ^{0.138} \rb.
  \label{eq:result_sr_t}
\end{equation}
Both equations perform reasonable well, leading to $R^2=0.93$ for the equivalent sand-grain roughness and $R^2=0.84$ for the temperature roughness function correlation. Similarly, both correlations can be tested on the corresponding external dataset $\cE$ as already performed during the neural network evaluation. Therefore, the coefficients are refitted to achieve a comparison of the form of the correlation rather than a comparison of the coefficients. This results in $R^2=0.95$ for the equivalent sand-grain correlation. The correlation for the temperature roughness function has a severely diminished predictive power for $\Delta \Theta^+$ of the external dataset, as visible in the $R^2=0.26$ value and in \cref{fig:compare_temperature}. This shortcoming might be attributed to several factors. First, the size of the dataset is limited and hence complicating a well-founded conclusion and second, the external dataset still includes roughness types never seen by the correlation before. In addition to these limitations, the temperature roughness function itself is subject to general restrictions due to the confined database (single Prandtl-number and $Re_\tau \approx 800$) as already elaborated.
It should be mentioned, that the evaluation of the temperature roughness function is performed on the true values of the equivalent sand-grain roughness. Hence, the presented results provide a best case scenario. For example, if the roughness correlation for $k_\text{s}$ (tuned given the correct values) is used and the refit performed on these predicted values, the coefficient of determination is slightly diminished ($R^2=0.83$). 

The great benefit of a simplistic roughness correlation is the interpretability of the equation compared to the complex form of the neural network. When analyzing the suggested correlation for $k_{\text{s}}$ the utilization of a measure of height ($k_{\text{99}}$), a measure of frontal area ($ES_x$) and a measure of coverage ($Sk$) is clearly visible. These three parameters are known to be elementary building blocks of predictive roughness correlation \cite{flack_2022a} and their importance is indicated through the presented data-driven approach as well.

Comparing the evolved equation in \cref{eq:result_sr_k} with already developed correlations (a selection is shown in \cref{tab:example-correlation}) show important similarities, besides the roughness parameters. As already noted, similar building blocks are present, namely $(Sk+a)^b$ or the direct scaling of $k_{\text{s}}$ using a measure of height. One important difference is the utilization of the effective slope $ES_x$ in the suggested correlation. Besides the correlation by \citet{forooghi_2017} and \citet{chan_2015} none of the presented correlation uses this parameter and hence the question arises, whether the integration is favorable. 

\begin{figure*}
  \centering
  \ifpreprint
    \hspace*{-2.2cm}%
  \fi
  \input{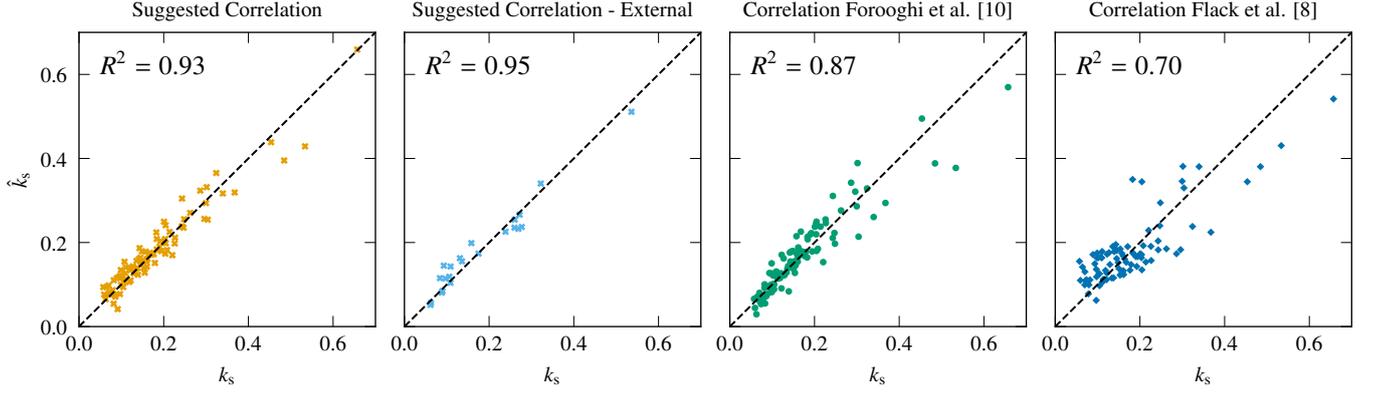}
  \caption{Comparison between true equivalent sand-grain roughness $k_{\text{s}}$ and the predicted value $\hat{k}_{\text{s}}$ of different roughness correlations using the database of \citet{yang_2023}. The suggested correlation was tested on the external dataset $\cE_k$ as well.}
  \label{fig:compare_correlation}
\end{figure*}
To further test the suggested correlation as well as selected literature correlation, all simulation results of the database from \citet{yang_2023} are used. The correlation is evaluated for the roughness parameters and compared against the value derived from the DNS. The results are visualized in \cref{fig:compare_correlation} including the evaluation on the external dataset $\cE_k$. For further analysis, two of the roughness correlations from \cref{tab:example-correlation} are evaluated in similar fashion. For a fair comparison of the general form of the roughness correlation the model constants of the equation are refitted to the data, enabling a comparison of the correlation shape rather than an assessment of the model constants (which are furthermore derived from different data sources). 
The coefficient of determination $R^2$ visualizes the favorable predictive behavior of the correlation derived by symbolic regression and the spread around the optimal fit is limited, even for larger values of the equivalent sand-grain roughness. Similarly, the suggested correlation is able to capture the behavior in the external dataset reasonably well. The deviation from this optimal fit is higher for the other two correlations, which is reflected in the $R^2$ value as well. 
It is noteworthy that the correlations contain different statistical scalar roughness parameters and the different $R^2$ values again show the favorable behavior of including a measure for size (e.g., $k_\text{99}$), coverage (e.g., $Sk$) and frontal area (e.g., $ES_x$)  for roughness correlations (see \cite{flack_2022a} as well).
Effective slope is directly related to the frontal solidity ($ES_x \sim \lambda_f$ \cite{thakkar_2017}) and can be understood as a measure of wavelength \cite{chung_2021}. A low effective slope corresponds to long wavelengths, and higher values are associated to rough surfaces with steep ascents and descents, as well as short wavelengths.
Large wavelength have limited impact on the skin-friction drag \cite{barros_2018,yang_2023} and \citet{yang_2023} showed that a rough surface filtered with a roughness dependent wavenumber (determined through a machine learning approach) yields similar values in the drag behavior. 
This motivated the investigation on the effect of wavelengths in the derived correlation via the effective slope.
\begin{figure}
  \centering
  \import{figures}{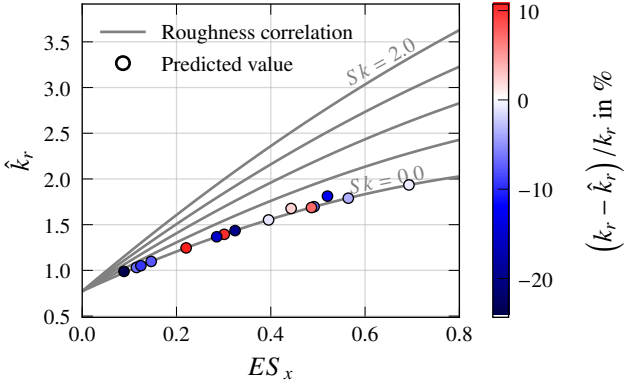}
  \caption{Prediction by the roughness correlation for $k_\text{r}$ for a constant values of skewness. For non-skewed surfaces, simulation results are added by their value predicted through the correlation and  color-coded by the deviation to the true value of the simulation.}
  \label{fig:kr_trend}
\end{figure}
In \cref{fig:kr_trend} the derived correlation is evaluated for specific skewness values and values of effective slope, as this is easily possible for roughness correlations. The roughness correlation follows the trends observed by \citet{kuwata_2023}, namely higher values for larger skewness and a decrease of slope for increasing effective slope. The latter is less prominent, which might be due to the different normalization values. In addition, simulated surfaces with zero skewness are investigated and added to \cref{fig:kr_trend} using the corresponding value of the roughness correlation. The difference to the value obtained from the simulation is then encoded by the color. 
We observe in \cref{fig:kr_trend}, that for a non-skewed surface the error of the suggested predictive correlation increases as the effective slope decreases (increasing wavelength), indicating again the importance of wavelengths and pointing towards the scalar roughness parameter $ES_x$. 
Despite the presented importance of the effective slope and other roughness parameters, these statistical measures are subject of recent discussions as well due to their lack in containing information on flow physics, see e.g., \citet{meneveau_2024} or \citet{bruno_2024}. One key observation is the inability of current surface parameters to incorporate sheltering effects, which are known to be of importance from other references as well (e.g., \cite{yang_2022a}). This leads to the formulation of new parameters, e.g., the effective distribution $ED$ \cite{bruno_2024} or the wind-shade model \cite{meneveau_2024}. Including more complex parameters in the terminal set may enhance correlation; however, it also increases the complexity of determining roughness parameters and is therefore beyond the scope of this study. \\

Correlations for the temperature roughness functions primarily focus on a direct relationship between $\Delta \Theta^+$ and the equivalent sand-grain roughness \cite{kadivar_2024a}, e.g., via the roughness Stanton-number
\begin{equation}
  \Delta \Theta^+ = \frac{1}{\kappa_{\theta}} \ln \lb k_\text{s}^+ \rb + A_\theta - C - St_k^{-1}.
  \label{eq:delta-theta}
\end{equation}
Here the relationship $St_k \sim (\lambda_f)^n (k_\text{s}^+)^{-p} Pr^{-m}$ is widely used as summarized by \citet{kadivar_2024a} from various literature resources. The close connection to \cref{eq:delta-theta-g} should be mentioned, as $g = St_k^{-1} + 8.5 Pr_t$ \cite{kadivar_2024a} using the turbulent Prandtl-number $Pr_t$. The proposed equation does not directly follow this relation and furthermore includes more scalar roughness parameters. The missing Prandtl-number can be explained through the missing Prandtl-number variation in the original dataset. It is important to note, that the building block $\ln (k_\text{s})$ is present in some candidates of the symbolic regression approach, however these equations resulted in higher loss values and hence were discarded. This observation improves the confidence of the symbolic regression approach as well as the basic equation (\ref{eq:delta-theta}).
For the presented dataset, the correlations by \citet{kays_2007} and \citet{kader_1977} were not able to predict the temperature roughness function in a sufficient manner when tested on the DNS results (visible in \cref{fig:compare_temperature}). The suggested correlation for the temperature roughness function improves the predictions, but is not able to fully capture the trends in the dataset, especially higher values of $\Delta \Theta^+$. Nevertheless, including statistical roughness parameters improved the correlation, similar to \citet{kuwata_2024}. Here, especially the skewness of the roughness as well as an additional measure of height is beneficial. Especially the latter points towards the increased wetted surface area of the roughness, but the picture is less clear, than for the velocity counterpart. For example, \citet{kuwata_2024} reported an improved prediction of the temperature roughness function by including effective slope, even though this study just investigated sinusoidal rough surfaces.

\section{Exploration of the derived models}
\label{sec:results}

\begin{figure*}
  \centering
  \ifpreprint
    \hspace*{-2.2cm}%
  \fi
  \includegraphics[scale=1.0]{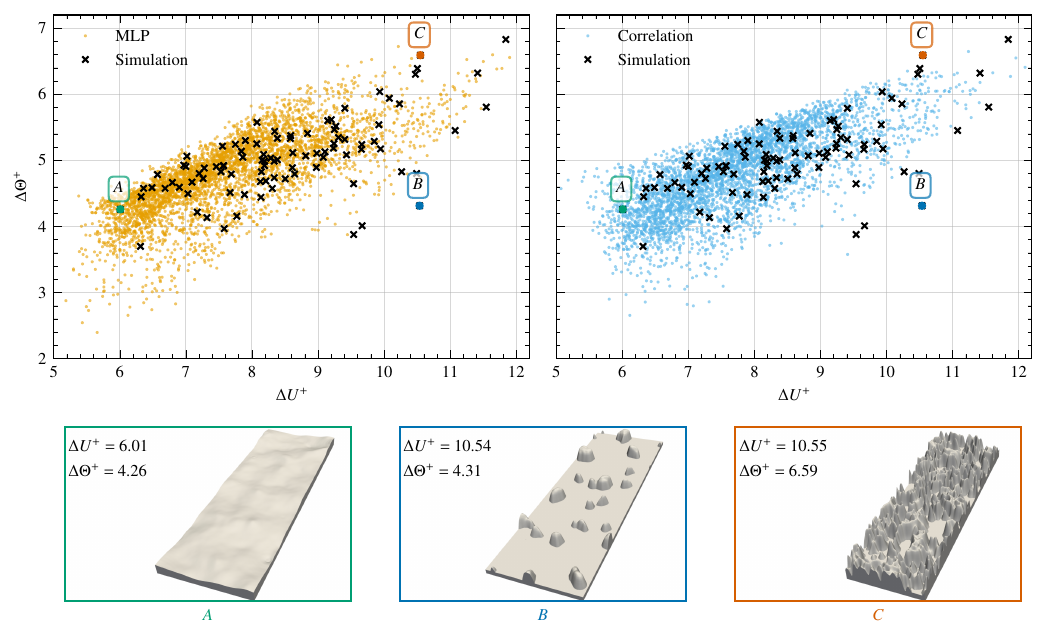}
  \caption{Evaluation of the temperature and velocity roughness function using the developed neural networks and the derived correlation for the dataset of $4200$ surfaces.}
  \label{fig:Exploration}
\end{figure*}
To explore the predictive capabilities of the neural network and the derived correlations, a prediction is generated for all $4200$ rough surfaces in a forward pass through the neural network.
The prediction based on the neural network as well as the correlation is visualized in \cref{fig:Exploration}. This visualization is enriched by the results of the direct numerical simulations of the original dataset as well, shown by the black cross symbol. In general, larger values of $\Delta U^+$ are typically accompanied by larger values of $\Delta \Theta^+$. However, it should be mentioned that this tendency exhibits a slope smaller than one and $\Delta \Theta^+$ is always smaller than $\Delta U^+$ for the investigated Prandtl-number $Pr=0.71$. 
A comparison of the neural network prediction and the prediction by the derived correlation serves as an assessment of the quality of the transfer from the network to the correlation. 
Especially in the region of higher values of $\Delta \Theta^+$ combined with low $\Delta U^+$ values, the correlation leads to predictions, whereas the neural networks do not, as seen in the missing datapoints in this region of \cref{fig:Exploration}. A similar behavior is observed at low values of $\Delta \Theta^+$ at higher $\Delta U^+$ value. A possible reason for these deviations can be the lower quality of the temperature roughness prediction described in \cref{sec:symbolic-regression}. In addition, in the evaluation of the $\Delta \Theta^+$ correlation in \cref{fig:Exploration} involves the predicted equivalent sand-grain roughness from the $k_\text{s}$-correlation, whereas in the symbolic regression algorithm the prediction by the neural network is used, introducing an additional source of error.
Nevertheless, the results of the direct numerical simulation align well with the results of the neural network and the roughness correlations. However, rough surfaces with low values of $\Delta \Theta^+$ at higher $\Delta U^+$ value are rather not captured well by the proposed correlations. Additionally, the number of simulations with low values of the temperature roughness function is limited, likely due to the active learning algorithm governed by the $k_\text{r}$ for the selection of the rough surfaces for simulation.

Furthermore, the DNS data shows data points, which do not follow the trend of the majority of rough surfaces. These examples are marked in \cref{fig:Exploration} using the labels $A$, $B$ and $C$. Surface $A$ and $B$ share similar values for the temperature roughness function, but there is a large disparity for the velocity roughness function. For roughness $B$ and $C$ this relation is reversed. 
The surface height profile of the three investigated surfaces is visualized in \cref{fig:Exploration} as well, clearly showing different characteristics. 
\begin{table}
  \centering
  \begin{tabular}{c c c c}
    \toprule
    Surface ID & $A$ & $B$ & $C$ \\
    \midrule
    $k_\text{t}$ & $0.0897$ & $0.1497$ & $0.1449$ \\
    $k_\text{rms}$ &  $0.0142$ & $0.0282$ & $0.0478$ \\ 
    $k_\text{avg}$ & $0.0421$ & $0.0087$ & $0.0552$ \\ 
    $Sk$ &  $0.1619$ & $3.4148$ & $0.2994$ \\ 
    $Ku$ & $3.0841$ & $13.788$ & $1.6953$ \\
    $ES_x$ & $0.0887$ & $0.1754$ & $1.5020$ \\
    \bottomrule
  \end{tabular}
  \caption{Statistical scalar parameters for the selected six roughness topographies (Note: $k$-values are normalized by channel half height $\delta$).}
  \label{tab:roughness_statistics}
\end{table}
These different characteristics are additionally observed in the scalar statistical parameters (see \cref{tab:roughness_statistics}). The surfaces with similar values in $\Delta U^+$ share values of the same ballpark for effective slope, whereas roughness $C$ has a significantly larger value of $ES_x$. Furthermore, $B$ has the highest skewness value and results in a higher value for $\Delta U^+$ but low $\Delta \Theta^+$.
The visualization indicates that a rough surface with a high value of a temperature roughness functions exhibits more frequent roughness height changes (low wavelength and hence larger wavenumbers), when compared to the other two surfaces. As effective slope can be understood as a measure of wavelength \cite{chung_2021} the statistical parameters underline this visual observation.
The drag behavior of rough surfaces depends on the wavelengths present in the roughness, wavenumbers below a roughness dependent wavenumber are drag-irrelevant \cite{yang_2023}. The importance of wavenumbers can be concluded from the analyzed roughness correlations as well, whereas the investigations of the temperature roughness function is less clear (see \cref{sec:symbolic-regression}). 
For uniform sinusoidal rough surfaces with a single length scale, \citet{zhong_2023} proposed a correlation for $\Delta \Theta^+$ depending solely on $Pr$ and $k_\text{s}$. Such a relation is not observed in the dataset used throughout this study, likely due to the irregular nature of the artificial rough surfaces investigated. 
The visual observation as well as the analysis of temperature roughness correlations indicate, that the temperature roughness function depends on the wavelengths of the rough surface, similar to the equivalent sand-grain roughness. This is further supported by the observation of \citet{rowin_2024} on sinusoidal surfaces with varying wavelengths. There, the Reynolds analogy breakdown depends on the roughness wavelength. Similarly, \citet{kuwata_2024} investigated the effect of steepness using sinusoidal rough surface. Here, the importance of effective slope as a measure of steepness was concluded. This aligns with the observation made by exploring the neural network, even though effective slope was not included in the expression evolved through symbolic regression.
Hence, additional investigations on irregular rough surfaces with filtered and non-filtered roughness height maps (similar to Refs. \cite{barros_2018,yang_2023}) for the temperature roughness function might be beneficial to broaden the understanding.
Additionally, the rough surface for high $\Delta \Theta^+$ is characterized by deep and often reoccurring valleys. In these sheltered cavities, recirculation zones can occur, and this phenomenon is observed in the literature as well \cite{peeters_2019,kuwata_2021}. The recirculation zones form a clear difference between velocity and temperature field. \citet{zhong_2023} suggested the ratio between attached and detached flow as a metric for the heat transfer. A scalar measure including this knowledge might be beneficial for the temperature roughness prediction, similar to the wind-shade model \cite{meneveau_2024}. \citet{rowin_2024} investigated this idea for sinusoidal rough surfaces, but a transfer to irregular rough surfaces is yet to be conducted, but based on literature source and the conducted study the scalar parameter effective slope or adjusted modifications might provide a viable route.

\section{Conclusion}
\label{sec:Conclusion}
The extensive database by \citet{yang_2023} is utilized to develop predictive tools for the velocity and temperature roughness function, representing the influence of rough surfaces. For this purpose, a neural network for the normalized equivalent sand-grain roughness $k_\text{r}$ and a network for the temperature roughness function $\Delta \Theta^+$ is developed. Both networks build on the power spectrum and the probability density function to predict the quantities of interest. Training is performed in a supervised learning environment created through the DNS results by \citet{yang_2023}. Here, temperature is regarded as a passive scalar. The roughness functions $\Delta U^+$ (and hence the normalized equivalent sand-grain roughness $k_\text{r}$) and $\Delta \Theta^+$ are computed from the temporal and spatial averaged profiles following the procedure by \citet{yang_2023}. 
Both networks provide a suitable approximation and decent predictive capabilities, indicated by a testing error of $\text{MAPE} = 13.4 \%$  and $\text{MAPE} = 4.3 \%$  regarding the normalized equivalent sand-grain roughness $k_\text{r}$ and temperature roughness function $\Delta \Theta^+$, respectively. The generalization capabilities on the external datasets is diminished due to the nature of different rough surfaces present. Furthermore, especially the prediction of $\Delta \Theta^+$ is limited to a single Prandtl-number $Pr=0.71$ and a constant friction Reynolds-number $Re_\tau \approx 800$, as no asymptotic or scaling behavior is known for passive scalars yet. Such a behavior is available for the velocity part, namely the fully-rough asymptote opens the possibility for a single length-scale in the fully rough regime, and hence the prediction of the normalized equivalent sand-grain roughness is more sophisticated and generalizable. \\

The developed neural networks serve as acceptable predictive tools, but are rather limited in understanding and explaining their approximation. Therefore, we employ a symbolic regression workflow to distill an interpretable and easy to apply roughness correlation from the data-driven networks. Based on the prediction of the networks for $4200$ artificial generated rough surfaces, an evolutionary algorithm is implemented~\cite{cranmer_2023} evolving mathematical expressions inspired by biological evolution. The statistical scalar roughness parameters (e.g., Skewness $Sk$ or average height $k_\text{avg}$) form the basic building blocks of the correlations and the evolutionary algorithm iteratively adjusts the population of equations to optimize the fitness function and complexity of the expression.
This procedure yields a robust correlation set, with the correlation for the equivalent sand-grain roughness exceeding other literature correlations, achieving a coefficient of determination of $R^2 = 0.93$. 
The predictive correlation for $\Delta \Theta^+$ is constrained by the limitations of the dataset. However, compared to other correlations (see \cite{kadivar_2024a}), it suggests that incorporating additional roughness parameters could enhance its predictive capabilities, similar to the conclusion by \citet{kuwata_2024}.
The derived correlations provide the possibility to investigate the influences of roughness shape on thermohydraulic flow properties. 
The distillation procedure confirms that this correlation should include a measure of height, slope, and coverage to predict the velocity roughness function, similar to already existing literature sources. 
Together with the evaluation of the neural network the importance of wavelength, shown through the statistical parameter $ES_x$, become apparent. 

In conclusion, the functional form derived through symbolic regression from the neural network provides a reasonable approximation and satisfactory generalization. 
Its simplistic structure, combined with basic statistical scalar parameters, serves as a streamlined surrogate for a more complex data-driven approach. 
Moreover, the derived correlations enhance the interpretability of the neural network predictions, offering valuable insights, namely the effect of roughness wavelength on the velocity and temperature roughness function. This insight opens the possibility for focused investigations in the future.

\section*{CRediT authorship contribution statement}
\textbf{S.D.:} Methodology, Software, Formal analysis, Investigation, Data curation,  Visualization, Writing – original draft
\textbf{J.Y.:} Methodology, Software, Investigation, Data curation, Writing - review \& editing
\textbf{P.F.:} Supervision, Writing - review \& editing
\textbf{B.F.:} Resources, Supervision, Funding acquisition, Writing - review \& editing
\textbf{A.S.:} Conceptualization, Supervision, Funding acquisition, Resources, Project administration, Writing - review \& editing

\section*{Acknowledgement}
This work was performed with the help of computing time on the  high-performance computer HoreKa at the NHR@KIT and the Large Scale Data Facility funded by the Ministry of Science, Research and the Arts Baden-Württemberg and by the Federal Ministry of Education and Research.

\section*{Funding sources}
S.D. gratefully acknowledges partial financial support from the Friedrich und Elisabeth Boysen-Foundation (Grant: BOY-192).

\section*{Declaration of competing interest}
The authors declare that they have no known competing financial interests or personal relationships that could have appeared to influence the work reported in this paper.

\section*{Data availability}
The data and scripts used in this study will be available on the author’s GitHub-webpage. DNS results for velocity are already available at \url{www.roughnessdatabase.org}.

\appendix
\section{Scalar roughness parameters}
\label{app:database}
In the following, the most important scalar roughness parameters are presented, used throughout this study, namely the meltdown or average height $k_\text{avg}$, root-mean-squared height $k_\text{rms}$, peak-to-trough height $k_\text{t}$, arithmetic average height $k_\text{a}$, skewness $Sk$, kurtosis $Ku$, effective slope $ES_x$ and the root-mean-squared slope $\xi_\text{rms}$. These can be computed using the given equations:
\begin{align}
  k_\text{avg} &= \la h(x_1,x_3) \ra \\
  k_\text{rms} &= \sqrt{\la h(x_1,x_3) - k_\text{avg})^2 \ra} \\
  k_\text{t} & = \max(h(x_1,x_3)) - \min(h(x_1,x_3))\\
  k_\text{a} & = \la \abs{h(x_1,x_3) - k_\text{avg}} \ra \\
  Sk & = \frac{\la (h(x_1,x_3) - k_\text{avg})^3 \ra}{k_{rms}^3}\\
  Ku & = \frac{\la (h(x_1,x_3) - k_\text{avg})^4 \ra}{k_{rms}^34}\\
  ES_x &= \Big\la \absg{\pd{h(x_1,x_3)}{x}} \Big\ra \\
  \xi_\text{rms} & = \sqrt{\la (\nabla h(x_1, x_3))^2 \ra}
\end{align}
It should be noted again, that the roughness height map is normalized using the channel half height $\delta$, and this propagates into dimension-related statistical variables.
These definitions follow \citet{chung_2021} and the operation $\la \cdot \ra$ denotes the average over the nominal surface $A$
\begin{equation}
  \la f(x,z) \ra = \frac{1}{A} \int_{A} f(x,z) \d A .
\end{equation}
Additionally the following scalar roughness parameters are used in this study:
\begin{itemize}
  \item $\bar{k}_\text{t}$: Mean subdivided peak-to-valley height
  \item $V_\text{mat}$: Integration of the cumulative distribution function of the height map normalized by the root-mean-squared height
  \item $AC_\text{len}$: Autocorrelation length for a two-dimensional surface scan following the computation of DIN EN ISO 25178-2 normalized using the rms-height
  \item $AC_x$: Autocorrelation length in $x$-direction computed as the mean of the correlations lengths for profile slices in $x$-direction and normalized using the rms-height
\end{itemize}

\begin{figure*}
  \centering
  \input{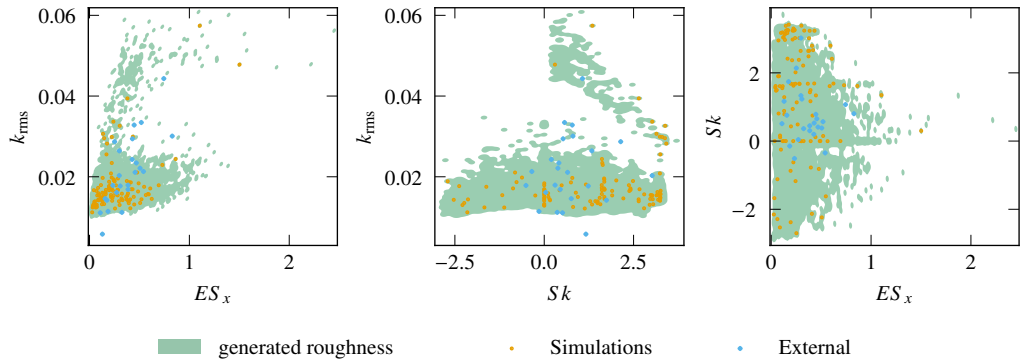}
  \caption{Distribution of selected statistical roughness parameters in the database of $4200$ rough surfaces as well as the simulated rough surfaces and the external dataset.}
  \label{fig:Histograms}
\end{figure*}
Additionally, an overview on the appearance of these quantities in the original database of \citet{yang_2023}, rough surfaces of the external dataset as well as simulated surfaces, are shown for reference in \cref{fig:Histograms}.

\bibliographystyle{elsarticle-num-names}
\bibliography{ref_clean, add_references}
\end{document}

%% file: macros.tex
\newcommand{\Ds}{\displaystyle}
\newcommand{\Ts}{\textstyle}
\newcommand{\Ss}{\scriptstyle}%
\renewcommand{\d}{{\,\rm  d}}
\newcommand{\pd}[2]{\displaystyle\frac{\partial #1}{\partial #2}}
\newcommand{\td}[2]{\frac{{\rm d} #1}{{\rm d} #2}}
\newcommand{\pdm}[4]{\displaystyle\frac{\partial^{#1} #2}{\partial #3 \partial #4}}
\newcommand{\abs}[1]{|#1|}
\newcommand{\absg}[1]{\left| #1 \right|}
\newcommand{\lb}{\left(}
\newcommand{\rb}{\right)}
\newcommand{\la}{\langle}
\newcommand{\ra}{\rangle}
\newcommand{\cred }{\color{red}}
\newcommand{\cblue }{\color{blue}}
\newcommand{\cgreen }{\color{green}}
\newcommand{\sty}[1]{\mbox{\boldmath $#1$}}
\newcommand{\styy}[1]{{\mathbb{#1}}}
\newcommand{\fe}{\sty{ e}}
\newcommand{\ff}{\sty{ f}}
\newcommand{\ffR}{\styy{ R}}
\newcommand{\cD}{{\cal D}}
\newcommand{\cE}{{\cal E}}
\newcommand{\cF}{{\cal F}}
\newcommand{\cM}{{\cal M}}
\newcommand{\cS}{{\cal S}}
\newcommand{\cT}{{\cal T}}
\newcommand{\cV}{{\cal V}}

%% file: figures/symbolic_regression.tex
\begin{tikzpicture}
    \node[database,label=left:$\cS$,database radius=0.75cm,database segment height=0.35cm] (data) at (0,0) {};

    \draw[rounded corners=5, fill=gray] (2.25,2) rectangle (2.75,5) node[pos=.5, rotate=-90, text=white] {PS, PDF, ...};
    \node [trapezium, minimum width=3cm,  minimum height=2cm, trapezium left angle=70, trapezium right angle=70, trapezium stretches, draw, rotate=-90, rounded corners=2] (A) at (4,3.5) {\rotatebox{90}{MLP}};
    \draw[rounded corners=5, fill=gray] (5.25,3) rectangle (5.75,4) node[pos=.5, rotate=-90, text=white] (L) {$k_\text{r}$};

    \draw[rounded corners=5] (2.25, -3) rectangle (5.75,-4.5) node[pos=.5, align=center] (Roughness) {Compute roughness \\properties};

    \matrix[ampersand replacement=\&] at (11, 0){
        \node (head2) {Data for Symbolic Regression};  \& \node {}; \\
        \node (table2) [shape=rectangle, draw, rounded corners=4] {
            \begin{tabular}{c | c| c}
                PS, PDF, \dots & $Sk$, $ES$, $k_{\text{t}}$, \dots & $k_{\text{r}}$ \\
                \hline
                $\vdots$ & $\vdots$ & $\vdots$ \\
            \end{tabular}
        }; \&
        \node {}; \\
    };

    \draw[thick, rounded corners=5] (15,4.5) rectangle (22.75,-3.35);
    \node[right] at (15,4.75) {\textbf{Symbolic Regression}};

    \draw[rounded corners=5] (15.25,2.2) rectangle (16,-3.1) node[pos=.5, rotate=-90] (P) {Populations};

    \draw (18, 1) rectangle (22.5,-0.5) node[pos=.5, align=left] (A) {Altering \\ $\rightarrow$ Mutation, Crossover, \\ \hspace{10pt} Simplification, Optimize};
    \draw (18, -1) rectangle (22.5, -1.75) node[pos=.5,xshift=-0.6cm, align=left] (E) {Evaluate with data};

    \draw[-latex] (17,0) arc[start angle=90, end angle=-180, radius=0.5cm];
    \draw[rounded corners=5] (15.25,3.5) rectangle (18.25, 4.25) node[pos=.5] (T) {Terminal set $\cM$};
    \draw[rounded corners=5] (18.75,3.5) rectangle (21.75, 4.25) node[pos=.5] (F) {Function set $\cF$};

    \draw[-latex, bend left] ([yshift=0.25cm]data.north) to (2.0,3.5);
    \draw[-latex, bend left] ([xshift=0.25cm]L.north) to [out=50, in=245]([xshift=-0.25cm]table2.west);
    \draw[-latex, bend right] ([xshift=0.5cm]Roughness.east) to [out=-40, in=115]([xshift=-0.25cm]table2.west);
    \draw[-latex, bend right] ([yshift=-0.25cm]data.south) to [out=-20, in=210]([xshift=-0.5cm]Roughness.west);
    \draw[-latex] ([xshift=0.25cm]table2.east) to (14.75,-0.27);
    \draw[-latex, bend left] (18.5, -1.9) to [out=30, in=160] (16.25, -2.9);
    \draw[-latex] (18.5, -0.6) to (18.5, -0.9);
    \draw[-latex, bend right] (16.25, 2.1)to [out=30, in=160] (18.5, 1.1);
    \draw[-latex] (15.60, 3.4) to (15.60, 2.3);
    \draw (19.1,3.4) -- (19.1, 3) -- (15.60, 3);

    \node at (18.875, -3.75) {\rotatebox{-90}{$\Rightarrow$}};
    \node[text width=3cm, align=center] at (18.875, -4.5) {Correlation $f_{\text{SR}}: \ffR^n \rightarrow \ffR$};

    \draw [decorate,decoration={brace,amplitude=10pt,mirror,raise=4pt},yshift=0pt]
    (2.25,1.75) -- (5.75,1.75) node [black,midway,yshift=-0.8cm] {$f_{\text{MLP}}:\ffR^m \rightarrow \ffR$};

\end{tikzpicture}

%% file: main.bbl
\begin{thebibliography}{55}
\expandafter\ifx\csname natexlab\endcsname\relax\def\natexlab#1{#1}\fi
\providecommand{\url}[1]{\texttt{#1}}
\providecommand{\href}[2]{#2}
\providecommand{\path}[1]{#1}
\providecommand{\DOIprefix}{doi:}
\providecommand{\ArXivprefix}{arXiv:}
\providecommand{\URLprefix}{URL: }
\providecommand{\Pubmedprefix}{pmid:}
\providecommand{\doi}[1]{\href{http://dx.doi.org/#1}{\path{#1}}}
\providecommand{\Pubmed}[1]{\href{pmid:#1}{\path{#1}}}
\providecommand{\bibinfo}[2]{#2}
\ifx\xfnm\relax \def\xfnm[#1]{\unskip,\space#1}\fi
\bibitem[{Yang et~al.(2023)Yang, Stroh, Lee, Bagheri, Frohnapfel, and
  Forooghi}]{yang_2023}
\bibinfo{author}{J.~Yang}, \bibinfo{author}{A.~Stroh},
  \bibinfo{author}{S.~Lee}, \bibinfo{author}{S.~Bagheri},
  \bibinfo{author}{B.~Frohnapfel}, \bibinfo{author}{P.~Forooghi},
\newblock \bibinfo{title}{Prediction of equivalent sand-grain size and
  identification of drag-relevant scales of roughness – a data-driven
  approach},
\newblock \bibinfo{journal}{J. Fluid Mech.} \bibinfo{volume}{975}
  (\bibinfo{year}{2023}) \bibinfo{pages}{A34}.
  \DOIprefix\doi{10.1017/jfm.2023.881}.
\bibitem[{Garg et~al.(2024)Garg, Sahut, Tuneskog, Nogenmyr, and
  Fureby}]{garg_2024}
\bibinfo{author}{H.~Garg}, \bibinfo{author}{G.~Sahut},
  \bibinfo{author}{E.~Tuneskog}, \bibinfo{author}{K.-J. Nogenmyr},
  \bibinfo{author}{C.~Fureby},
\newblock \bibinfo{title}{Large eddy simulations of flow over additively
  manufactured surfaces: {Impact} of roughness and skewness on turbulent heat
  transfer},
\newblock \bibinfo{journal}{Phys. Fluids} \bibinfo{volume}{36}
  (\bibinfo{year}{2024}) \bibinfo{pages}{085143}.
  \DOIprefix\doi{10.1063/5.0221006}.
\bibitem[{Bons(2010)}]{bons_2010}
\bibinfo{author}{J.~P. Bons},
\newblock \bibinfo{title}{A {Review} of {Surface} {Roughness} {Effects} in
  {Gas} {Turbines}},
\newblock \bibinfo{journal}{J. Turbomach.} \bibinfo{volume}{132}
  (\bibinfo{year}{2010}). \DOIprefix\doi{10.1115/1.3066315}.
\bibitem[{Hama(1954)}]{hama_1954}
\bibinfo{author}{F.~R. Hama},
\newblock \bibinfo{title}{Boundary-layer characteristics for smooth and rough
  surfaces},
\newblock \bibinfo{journal}{Society of Naval Architects and Marine Engineers}
  (\bibinfo{year}{1954}).
\bibitem[{Clauser(1954)}]{clauser_1954}
\bibinfo{author}{F.~H. Clauser},
\newblock \bibinfo{title}{Turbulent {Boundary} {Layers} in {Adverse} {Pressure}
  {Gradients}},
\newblock \bibinfo{journal}{J. Astronaut. Sci.} \bibinfo{volume}{21}
  (\bibinfo{year}{1954}) \bibinfo{pages}{91--108}.
  \DOIprefix\doi{10.2514/8.2938}.
\bibitem[{Nikuradse(1933)}]{nikuradse_1933}
\bibinfo{author}{J.~Nikuradse},
\newblock \bibinfo{title}{Strömungsgesetz in {Rauhen} {Rohren}},
\newblock \bibinfo{journal}{VDI-Forschungsheft} \bibinfo{volume}{361:1}
  (\bibinfo{year}{1933}).
\bibitem[{Abdelaziz et~al.(2024)Abdelaziz, Djenidi, Ghayesh, and
  Chin}]{abdelaziz_2024}
\bibinfo{author}{M.~Abdelaziz}, \bibinfo{author}{L.~Djenidi},
  \bibinfo{author}{M.~H. Ghayesh}, \bibinfo{author}{R.~Chin},
\newblock \bibinfo{title}{On predictive models for the equivalent sand grain
  roughness for wall-bounded turbulent flows},
\newblock \bibinfo{journal}{Phys. Fluids} \bibinfo{volume}{36}
  (\bibinfo{year}{2024}) \bibinfo{pages}{015125}.
  \DOIprefix\doi{10.1063/5.0178798}.
\bibitem[{Flack et~al.(2016)Flack, Schultz, Barros, and Kim}]{flack_2016}
\bibinfo{author}{K.~Flack}, \bibinfo{author}{M.~Schultz},
  \bibinfo{author}{J.~Barros}, \bibinfo{author}{Y.~Kim},
\newblock \bibinfo{title}{Skin-friction behavior in the transitionally-rough
  regime},
\newblock \bibinfo{journal}{Int. J. Heat Fluid Flow} \bibinfo{volume}{61}
  (\bibinfo{year}{2016}) \bibinfo{pages}{21--30}.
  \DOIprefix\doi{10.1016/j.ijheatfluidflow.2016.05.008}.
\bibitem[{Flack et~al.(2020)Flack, Schultz, and Barros}]{flack_2020}
\bibinfo{author}{K.~A. Flack}, \bibinfo{author}{M.~P. Schultz},
  \bibinfo{author}{J.~M. Barros},
\newblock \bibinfo{title}{Skin {Friction} {Measurements} of
  {Systematically}-{Varied} {Roughness}: {Probing} the {Role} of {Roughness}
  {Amplitude} and {Skewness}},
\newblock \bibinfo{journal}{Flow Turbul. Combust.} \bibinfo{volume}{104}
  (\bibinfo{year}{2020}) \bibinfo{pages}{317--329}.
  \DOIprefix\doi{10.1007/s10494-019-00077-1}.
\bibitem[{Forooghi et~al.(2017)Forooghi, Stroh, Magagnato, Jakirlić, and
  Frohnapfel}]{forooghi_2017}
\bibinfo{author}{P.~Forooghi}, \bibinfo{author}{A.~Stroh},
  \bibinfo{author}{F.~Magagnato}, \bibinfo{author}{S.~Jakirlić},
  \bibinfo{author}{B.~Frohnapfel},
\newblock \bibinfo{title}{Toward a {Universal} {Roughness} {Correlation}},
\newblock \bibinfo{journal}{J. Fluids Eng.} \bibinfo{volume}{139}
  (\bibinfo{year}{2017}) \bibinfo{pages}{121201}.
  \DOIprefix\doi{10.1115/1.4037280}.
\bibitem[{Kuwata and Kawaguchi(2019)}]{kuwata_2019}
\bibinfo{author}{Y.~Kuwata}, \bibinfo{author}{Y.~Kawaguchi},
\newblock \bibinfo{title}{Direct numerical simulation of turbulence over
  resolved and modeled rough walls with irregularly distributed roughness},
\newblock \bibinfo{journal}{Int. J. Heat Fluid Flow} \bibinfo{volume}{77}
  (\bibinfo{year}{2019}) \bibinfo{pages}{1--18}.
  \DOIprefix\doi{10.1016/j.ijheatfluidflow.2019.02.009}.
\bibitem[{Bons(2002)}]{bons_2002}
\bibinfo{author}{J.~P. Bons},
\newblock \bibinfo{title}{St and cf {Augmentation} for {Real} {Turbine}
  {Roughness} {With} {Elevated} {Freestream} {Turbulence}},
\newblock \bibinfo{journal}{J. Turbomach.} \bibinfo{volume}{124}
  (\bibinfo{year}{2002}) \bibinfo{pages}{632--644}.
  \DOIprefix\doi{10.1115/1.1505851}.
\bibitem[{Chan et~al.(2015)Chan, MacDonald, Chung, Hutchins, and
  Ooi}]{chan_2015}
\bibinfo{author}{L.~Chan}, \bibinfo{author}{M.~MacDonald},
  \bibinfo{author}{D.~Chung}, \bibinfo{author}{N.~Hutchins},
  \bibinfo{author}{A.~Ooi},
\newblock \bibinfo{title}{A systematic investigation of roughness height and
  wavelength in turbulent pipe flow in the transitionally rough regime},
\newblock \bibinfo{journal}{J. Fluid Mech.} \bibinfo{volume}{771}
  (\bibinfo{year}{2015}) \bibinfo{pages}{743--777}.
  \DOIprefix\doi{10.1017/jfm.2015.172}.
\bibitem[{Chung et~al.(2021)Chung, Hutchins, Schultz, and Flack}]{chung_2021}
\bibinfo{author}{D.~Chung}, \bibinfo{author}{N.~Hutchins},
  \bibinfo{author}{M.~P. Schultz}, \bibinfo{author}{K.~A. Flack},
\newblock \bibinfo{title}{Predicting the {Drag} of {Rough} {Surfaces}},
\newblock \bibinfo{journal}{Annu. Rev. Fluid Mech.} \bibinfo{volume}{53}
  (\bibinfo{year}{2021}) \bibinfo{pages}{439--471}.
  \DOIprefix\doi{10.1146/annurev-fluid-062520-115127}.
\bibitem[{Flack and Chung(2022)}]{flack_2022a}
\bibinfo{author}{K.~A. Flack}, \bibinfo{author}{D.~Chung},
\newblock \bibinfo{title}{Important {Parameters} for a {Predictive} {Model} of
  ks for {Zero}-{Pressure}-{Gradient} {Flows}},
\newblock \bibinfo{journal}{AIAA J.} \bibinfo{volume}{60}
  (\bibinfo{year}{2022}) \bibinfo{pages}{5923--5931}.
  \DOIprefix\doi{10.2514/1.J061891}.
\bibitem[{Jouybari et~al.(2021)Jouybari, Yuan, Brereton, and
  Murillo}]{jouybari_2021}
\bibinfo{author}{M.~A. Jouybari}, \bibinfo{author}{J.~Yuan},
  \bibinfo{author}{G.~J. Brereton}, \bibinfo{author}{M.~S. Murillo},
\newblock \bibinfo{title}{Data-driven prediction of the equivalent sand-grain
  height in rough-wall turbulent flows},
\newblock \bibinfo{journal}{J. Fluid Mech.} \bibinfo{volume}{912}
  (\bibinfo{year}{2021}) \bibinfo{pages}{A8}.
  \DOIprefix\doi{10.1017/jfm.2020.1085}.
\bibitem[{Sanhueza et~al.(2023)Sanhueza, Akkerman, and
  Peeters}]{sanhueza_2023a}
\bibinfo{author}{R.~D. Sanhueza}, \bibinfo{author}{I.~Akkerman},
  \bibinfo{author}{J.~W. Peeters},
\newblock \bibinfo{title}{Machine learning for the prediction of the local skin
  friction factors and {Nusselt} numbers in turbulent flows past rough
  surfaces},
\newblock \bibinfo{journal}{Int. J. Heat Fluid Flow} \bibinfo{volume}{103}
  (\bibinfo{year}{2023}) \bibinfo{pages}{109204}.
  \DOIprefix\doi{10.1016/j.ijheatfluidflow.2023.109204}.
\bibitem[{Kadivar and Garg(2024)}]{kadivar_2024a}
\bibinfo{author}{M.~Kadivar}, \bibinfo{author}{H.~Garg},
\newblock \bibinfo{title}{Turbulent {Heat} {Transfer} over roughness: a
  comprehensive review of theories and turbulent flow structure},
\newblock \bibinfo{journal}{Int. J. Thermofluids}  (\bibinfo{year}{2024})
  \bibinfo{pages}{100967}. \DOIprefix\doi{10.1016/j.ijft.2024.100967}.
\bibitem[{Dipprey and Sabersky(1963)}]{dipprey_1963}
\bibinfo{author}{D.~F. Dipprey}, \bibinfo{author}{R.~H. Sabersky},
\newblock \bibinfo{title}{Heat and momentum transfer in smooth and rough tubes
  at various prandtl numbers},
\newblock \bibinfo{journal}{Int. J. Heat Mass Transfer} \bibinfo{volume}{6}
  (\bibinfo{year}{1963}) \bibinfo{pages}{329--353}.
  \DOIprefix\doi{10.1016/0017-9310(63)90097-8}.
\bibitem[{Kader(1981)}]{kader_1981}
\bibinfo{author}{B.~Kader},
\newblock \bibinfo{title}{Temperature and concentration profiles in fully
  turbulent boundary layers},
\newblock \bibinfo{journal}{Int. J. Heat Mass Transfer} \bibinfo{volume}{24}
  (\bibinfo{year}{1981}) \bibinfo{pages}{1541--1544}.
  \DOIprefix\doi{10.1016/0017-9310(81)90220-9}.
\bibitem[{Hantsis and Piomelli(2024)}]{hantsis_2024}
\bibinfo{author}{Z.~Hantsis}, \bibinfo{author}{U.~Piomelli},
\newblock \bibinfo{title}{Numerical {Simulations} of {Scalar} {Transport} on
  {Rough} {Surfaces}},
\newblock \bibinfo{journal}{Fluids} \bibinfo{volume}{9} (\bibinfo{year}{2024})
  \bibinfo{pages}{159}. \DOIprefix\doi{10.3390/fluids9070159}.
\bibitem[{Angelis et~al.(2023)Angelis, Sofos, and Karakasidis}]{angelis_2023}
\bibinfo{author}{D.~Angelis}, \bibinfo{author}{F.~Sofos},
  \bibinfo{author}{T.~E. Karakasidis},
\newblock \bibinfo{title}{Artificial {Intelligence} in {Physical} {Sciences}:
  {Symbolic} {Regression} {Trends} and {Perspectives}},
\newblock \bibinfo{journal}{Arch. Comput. Methods Eng.} \bibinfo{volume}{30}
  (\bibinfo{year}{2023}) \bibinfo{pages}{3845--3865}.
  \DOIprefix\doi{10.1007/s11831-023-09922-z}.
\bibitem[{Gerwin(1974)}]{gerwin_1974}
\bibinfo{author}{D.~Gerwin},
\newblock \bibinfo{title}{Information processing, data inferences, and
  scientific generalization},
\newblock \bibinfo{journal}{Behav. Sci.} \bibinfo{volume}{19}
  (\bibinfo{year}{1974}) \bibinfo{pages}{314--325}.
  \DOIprefix\doi{10.1002/bs.3830190504}.
\bibitem[{Langley(1981)}]{langley_1981}
\bibinfo{author}{P.~Langley},
\newblock \bibinfo{title}{Data-{Driven} {Discovery} of {Physical} {Laws}},
\newblock \bibinfo{journal}{Cogn. Sci.} \bibinfo{volume}{5}
  (\bibinfo{year}{1981}) \bibinfo{pages}{31--54}.
  \DOIprefix\doi{10.1111/j.1551-6708.1981.tb00869.x}.
\bibitem[{Koza(1994)}]{koza_1994}
\bibinfo{author}{J.~R. Koza},
\newblock \bibinfo{title}{Genetic programming as a means for programming
  computers by natural selection},
\newblock \bibinfo{journal}{Stat. Comput.} \bibinfo{volume}{4}
  (\bibinfo{year}{1994}) \bibinfo{pages}{87--112}.
  \DOIprefix\doi{10.1007/BF00175355}.
\bibitem[{Cranmer(2023)}]{cranmer_2023}
\bibinfo{author}{M.~Cranmer},
\newblock \bibinfo{title}{Interpretable {Machine} {Learning} for {Science} with
  {PySR} and {SymbolicRegression}.jl},
\newblock \bibinfo{journal}{arXiv}  (\bibinfo{year}{2023})
  \bibinfo{pages}{2305.01582}. \URLprefix
  \url{https://arxiv.org/abs/2305.01582}.
\bibitem[{Reuter et~al.(2022)Reuter, Cendrollu, Evrard, Mostaghim, and van
  Wachem}]{reuter_2022}
\bibinfo{author}{J.~Reuter}, \bibinfo{author}{M.~Cendrollu},
  \bibinfo{author}{F.~Evrard}, \bibinfo{author}{S.~Mostaghim},
  \bibinfo{author}{B.~van Wachem},
\newblock \bibinfo{title}{Towards {Improving} {Simulations} of {Flows} around
  {Spherical} {Particles} {Using} {Genetic} {Programming}},
\newblock in: \bibinfo{booktitle}{2022 {IEEE} {Congress} on {Evolutionary}
  {Computation} ({CEC})}, \bibinfo{year}{2022}, pp. \bibinfo{pages}{1--8}.
  \DOIprefix\doi{10.1109/CEC55065.2022.9870301}.
\bibitem[{Schmelzer et~al.(2020)Schmelzer, Dwight, and
  Cinnella}]{schmelzer_2020}
\bibinfo{author}{M.~Schmelzer}, \bibinfo{author}{R.~P. Dwight},
  \bibinfo{author}{P.~Cinnella},
\newblock \bibinfo{title}{Discovery of {Algebraic} {Reynolds}-{Stress} {Models}
  {Using} {Sparse} {Symbolic} {Regression}},
\newblock \bibinfo{journal}{Flow Turbul. Combust.} \bibinfo{volume}{104}
  (\bibinfo{year}{2020}) \bibinfo{pages}{579--603}.
  \DOIprefix\doi{10.1007/s10494-019-00089-x}.
\bibitem[{Pérez-Ràfols and Almqvist(2019)}]{perez-rafols_2019}
\bibinfo{author}{F.~Pérez-Ràfols}, \bibinfo{author}{A.~Almqvist},
\newblock \bibinfo{title}{Generating randomly rough surfaces with given height
  probability distribution and power spectrum},
\newblock \bibinfo{journal}{Tribol. Int.} \bibinfo{volume}{131}
  (\bibinfo{year}{2019}) \bibinfo{pages}{591--604}.
  \DOIprefix\doi{10.1016/j.triboint.2018.11.020}.
\bibitem[{Persson(2023)}]{persson_2023}
\bibinfo{author}{B.~N.~J. Persson},
\newblock \bibinfo{title}{On the {Use} of {Surface} {Roughness} {Parameters}},
\newblock \bibinfo{journal}{Tribol. Lett.} \bibinfo{volume}{71}
  (\bibinfo{year}{2023}) \bibinfo{pages}{29}.
  \DOIprefix\doi{10.1007/s11249-023-01700-z}.
\bibitem[{Yang et~al.(2022)Yang, Stroh, Chung, and Forooghi}]{yang_2022a}
\bibinfo{author}{J.~Yang}, \bibinfo{author}{A.~Stroh},
  \bibinfo{author}{D.~Chung}, \bibinfo{author}{P.~Forooghi},
\newblock \bibinfo{title}{Direct numerical simulation-based characterization of
  pseudo-random roughness in minimal channels},
\newblock \bibinfo{journal}{J. Fluid Mech.} \bibinfo{volume}{941}
  (\bibinfo{year}{2022}) \bibinfo{pages}{A47}.
  \DOIprefix\doi{10.1017/jfm.2022.331}.
\bibitem[{Chevalier et~al.(2007)Chevalier, Schlatter, Lundbladh, and
  Henningson}]{chevalier_2007}
\bibinfo{author}{M.~Chevalier}, \bibinfo{author}{P.~Schlatter},
  \bibinfo{author}{A.~Lundbladh}, \bibinfo{author}{D.~S. Henningson},
  \bibinfo{title}{{SIMSON} : {A} {Pseudo}-{Spectral} {Solver} for
  {Incompressible} {Boundary} {Layer} {Flows}}, \bibinfo{type}{Technical
  Report} \bibinfo{number}{Trita-MEK 2007:07}, KTH Mechanics,
  \bibinfo{year}{2007}.
\bibitem[{Goldstein et~al.(1993)Goldstein, Handler, and
  Sirovich}]{goldstein_1993}
\bibinfo{author}{D.~Goldstein}, \bibinfo{author}{R.~Handler},
  \bibinfo{author}{L.~Sirovich},
\newblock \bibinfo{title}{Modeling a {No}-{Slip} {Flow} {Boundary} with an
  {External} {Force} {Field}},
\newblock \bibinfo{journal}{J. Comput. Phys.} \bibinfo{volume}{105}
  (\bibinfo{year}{1993}) \bibinfo{pages}{354--366}.
  \DOIprefix\doi{10.1006/jcph.1993.1081}.
\bibitem[{Jiménez and Moin(1991)}]{jimenez_1991}
\bibinfo{author}{J.~Jiménez}, \bibinfo{author}{P.~Moin},
\newblock \bibinfo{title}{The minimal flow unit in near-wall turbulence},
\newblock \bibinfo{journal}{J. Fluid Mech.} \bibinfo{volume}{225}
  (\bibinfo{year}{1991}) \bibinfo{pages}{213--240}.
  \DOIprefix\doi{10.1017/S0022112091002033}.
\bibitem[{Chung et~al.(2015)Chung, Chan, MacDonald, Hutchins, and
  Ooi}]{chung_2015}
\bibinfo{author}{D.~Chung}, \bibinfo{author}{L.~Chan},
  \bibinfo{author}{M.~MacDonald}, \bibinfo{author}{N.~Hutchins},
  \bibinfo{author}{A.~Ooi},
\newblock \bibinfo{title}{A fast direct numerical simulation method for
  characterising hydraulic roughness},
\newblock \bibinfo{journal}{J. Fluid Mech.} \bibinfo{volume}{773}
  (\bibinfo{year}{2015}) \bibinfo{pages}{418--431}.
  \DOIprefix\doi{10.1017/jfm.2015.230}.
\bibitem[{MacDonald et~al.(2016)MacDonald, Chung, Hutchins, Chan, Ooi, and
  García-Mayoral}]{macdonald_2016}
\bibinfo{author}{M.~MacDonald}, \bibinfo{author}{D.~Chung},
  \bibinfo{author}{N.~Hutchins}, \bibinfo{author}{L.~Chan},
  \bibinfo{author}{A.~Ooi}, \bibinfo{author}{R.~García-Mayoral},
\newblock \bibinfo{title}{The minimal channel: a fast and direct method for
  characterising roughness},
\newblock \bibinfo{journal}{J. Phys. Conf. Ser.} \bibinfo{volume}{708}
  (\bibinfo{year}{2016}) \bibinfo{pages}{012010}.
  \DOIprefix\doi{10.1088/1742-6596/708/1/012010}.
\bibitem[{Yang et~al.(2023)Yang, Velandia, Bansmer, Stroh, and
  Forooghi}]{yang_2023d}
\bibinfo{author}{J.~Yang}, \bibinfo{author}{J.~Velandia},
  \bibinfo{author}{S.~Bansmer}, \bibinfo{author}{A.~Stroh},
  \bibinfo{author}{P.~Forooghi},
\newblock \bibinfo{title}{A comparison of hydrodynamic and thermal properties
  of artificially generated against realistic rough surfaces},
\newblock \bibinfo{journal}{Int. J. Heat Fluid Flow} \bibinfo{volume}{99}
  (\bibinfo{year}{2023}) \bibinfo{pages}{109093}.
  \DOIprefix\doi{10.1016/j.ijheatfluidflow.2022.109093}.
\bibitem[{Jackson(1981)}]{jackson_1981}
\bibinfo{author}{P.~S. Jackson},
\newblock \bibinfo{title}{On the displacement height in the logarithmic
  velocity profile},
\newblock \bibinfo{journal}{J. Fluid Mech.} \bibinfo{volume}{111}
  (\bibinfo{year}{1981}) \bibinfo{pages}{15--25}.
  \DOIprefix\doi{10.1017/S0022112081002279}.
\bibitem[{Flack and Schultz(2010)}]{flack_2010}
\bibinfo{author}{K.~A. Flack}, \bibinfo{author}{M.~P. Schultz},
\newblock \bibinfo{title}{Review of {Hydraulic} {Roughness} {Scales} in the
  {Fully} {Rough} {Regime}},
\newblock \bibinfo{journal}{J. Fluids Eng.} \bibinfo{volume}{132}
  (\bibinfo{year}{2010}) \bibinfo{pages}{041203}.
  \DOIprefix\doi{10.1115/1.4001492}.
\bibitem[{Pedregosa et~al.(2011)Pedregosa, Varoquaux, Gramfort, Michel,
  Thirion, Grisel, Blondel, Prettenhofer, Weiss, Dubourg, Vanderplas, Passos,
  Cournapeau, Brucher, Perrot, and Duchesnay}]{pedregosa_2011}
\bibinfo{author}{F.~Pedregosa}, \bibinfo{author}{G.~Varoquaux},
  \bibinfo{author}{A.~Gramfort}, \bibinfo{author}{V.~Michel},
  \bibinfo{author}{B.~Thirion}, \bibinfo{author}{O.~Grisel},
  \bibinfo{author}{M.~Blondel}, \bibinfo{author}{P.~Prettenhofer},
  \bibinfo{author}{R.~Weiss}, \bibinfo{author}{V.~Dubourg},
  \bibinfo{author}{J.~Vanderplas}, \bibinfo{author}{A.~Passos},
  \bibinfo{author}{D.~Cournapeau}, \bibinfo{author}{M.~Brucher},
  \bibinfo{author}{M.~Perrot}, \bibinfo{author}{E.~Duchesnay},
\newblock \bibinfo{title}{Scikit-learn: {Machine} {Learning} in {Python}},
\newblock \bibinfo{journal}{J. Mach. Learn. Res.} \bibinfo{volume}{12}
  (\bibinfo{year}{2011}) \bibinfo{pages}{2825--2830}.
\bibitem[{Zhong et~al.(2023)Zhong, Hutchins, and Chung}]{zhong_2023}
\bibinfo{author}{K.~Zhong}, \bibinfo{author}{N.~Hutchins},
  \bibinfo{author}{D.~Chung},
\newblock \bibinfo{title}{Heat-transfer scaling at moderate {Prandtl} numbers
  in the fully rough regime},
\newblock \bibinfo{journal}{J. Fluid Mech.} \bibinfo{volume}{959}
  (\bibinfo{year}{2023}) \bibinfo{pages}{A8}.
  \DOIprefix\doi{10.1017/jfm.2023.125}.
\bibitem[{O'Malley et~al.(2019)O'Malley, Bursztein, Long, Chollet, Jin,
  Invernizzi et~al.}]{omalley_2019}
\bibinfo{author}{T.~O'Malley}, \bibinfo{author}{E.~Bursztein},
  \bibinfo{author}{J.~Long}, \bibinfo{author}{F.~Chollet},
  \bibinfo{author}{H.~Jin}, \bibinfo{author}{L.~Invernizzi}, et~al.,
  \bibinfo{title}{{KerasTuner}}, \bibinfo{year}{2019}.
\bibitem[{Kruse et~al.(2015)Kruse, Borgelt, Braune, Klawonn, Moewes, and
  Steinbrecher}]{kruse_2015}
\bibinfo{author}{R.~Kruse}, \bibinfo{author}{C.~Borgelt},
  \bibinfo{author}{C.~Braune}, \bibinfo{author}{F.~Klawonn},
  \bibinfo{author}{C.~Moewes}, \bibinfo{author}{M.~Steinbrecher},
  \bibinfo{title}{Computational {Intelligence}: {Eine} methodische
  {Einführung} in {Künstliche} {Neuronale} {Netze}, {Evolutionäre}
  {Algorithmen}, {Fuzzy}-{Systeme} und {Bayes}-{Netze}},
  \bibinfo{publisher}{Springer Fachmedien Wiesbaden},
  \bibinfo{address}{Wiesbaden}, \bibinfo{year}{2015}.
  \DOIprefix\doi{10.1007/978-3-658-10904-2}.
\bibitem[{Flack and Schultz(2022)}]{flack_2022}
\bibinfo{author}{K.~A. Flack}, \bibinfo{author}{M.~P. Schultz},
\newblock \bibinfo{title}{Hydraulic characterization of sandpaper roughness},
\newblock \bibinfo{journal}{Exp. Fluids} \bibinfo{volume}{64}
  (\bibinfo{year}{2022}) \bibinfo{pages}{3}.
  \DOIprefix\doi{10.1007/s00348-022-03544-0}.
\bibitem[{Kuwata et~al.(2023)Kuwata, Yamamoto, Tabata, and Suga}]{kuwata_2023}
\bibinfo{author}{Y.~Kuwata}, \bibinfo{author}{Y.~Yamamoto},
  \bibinfo{author}{S.~Tabata}, \bibinfo{author}{K.~Suga},
\newblock \bibinfo{title}{Scaling of the roughness effects in turbulent flows
  over systematically-varied irregular rough surfaces},
\newblock \bibinfo{journal}{Int. J. Heat Fluid Flow} \bibinfo{volume}{101}
  (\bibinfo{year}{2023}) \bibinfo{pages}{109130}.
  \DOIprefix\doi{10.1016/j.ijheatfluidflow.2023.109130}.
\bibitem[{Thakkar et~al.(2017)Thakkar, Busse, and Sandham}]{thakkar_2017}
\bibinfo{author}{M.~Thakkar}, \bibinfo{author}{A.~Busse},
  \bibinfo{author}{N.~Sandham},
\newblock \bibinfo{title}{Surface correlations of hydrodynamic drag for
  transitionally rough engineering surfaces},
\newblock \bibinfo{journal}{J. Turbul.} \bibinfo{volume}{18}
  (\bibinfo{year}{2017}) \bibinfo{pages}{138--169}.
  \DOIprefix\doi{10.1080/14685248.2016.1258119}.
\bibitem[{Barros et~al.(2018)Barros, Schultz, and Flack}]{barros_2018}
\bibinfo{author}{J.~M. Barros}, \bibinfo{author}{M.~P. Schultz},
  \bibinfo{author}{K.~A. Flack},
\newblock \bibinfo{title}{Measurements of skin-friction of systematically
  generated surface roughness},
\newblock \bibinfo{journal}{Int. J. Heat Fluid Flow} \bibinfo{volume}{72}
  (\bibinfo{year}{2018}) \bibinfo{pages}{1--7}.
  \DOIprefix\doi{10.1016/j.ijheatfluidflow.2018.04.015}.
\bibitem[{Meneveau et~al.(2024)Meneveau, Hutchins, and Chung}]{meneveau_2024}
\bibinfo{author}{C.~Meneveau}, \bibinfo{author}{N.~Hutchins},
  \bibinfo{author}{D.~Chung},
\newblock \bibinfo{title}{The wind-shade roughness model for turbulent
  wall-bounded flows},
\newblock \bibinfo{journal}{J. Fluid Mech.} \bibinfo{volume}{1001}
  (\bibinfo{year}{2024}) \bibinfo{pages}{A3}.
  \DOIprefix\doi{10.1017/jfm.2024.971}.
\bibitem[{Bruno et~al.(2024)Bruno, Leonardi, and Marchis}]{bruno_2024}
\bibinfo{author}{F.~Bruno}, \bibinfo{author}{S.~Leonardi},
  \bibinfo{author}{M.~D. Marchis},
\newblock \bibinfo{title}{Towards a new roughness parametrization through the
  effective distribution function},
\newblock \bibinfo{journal}{J. Fluid Mech.} \bibinfo{volume}{999}
  (\bibinfo{year}{2024}) \bibinfo{pages}{A26}.
  \DOIprefix\doi{10.1017/jfm.2024.970}.
\bibitem[{Kays and Crawford(2007)}]{kays_2007}
\bibinfo{author}{W.~M. Kays}, \bibinfo{author}{M.~E. Crawford},
  \bibinfo{title}{Convective heat and mass transfer}, {McGraw}-{Hill} series in
  mechanical engineering, \bibinfo{edition}{4th international} ed.,
  \bibinfo{publisher}{McGraw-Hill}, \bibinfo{address}{Boston},
  \bibinfo{year}{2007}.
\bibitem[{Kader and Yaglom(1977)}]{kader_1977}
\bibinfo{author}{B.~Kader}, \bibinfo{author}{A.~Yaglom},
\newblock \bibinfo{title}{Turbulent heat and mass transfer from a wall with
  parallel roughness ridges},
\newblock \bibinfo{journal}{Int. J. Heat Mass Transfer} \bibinfo{volume}{20}
  (\bibinfo{year}{1977}) \bibinfo{pages}{345--357}.
  \DOIprefix\doi{10.1016/0017-9310(77)90156-9}.
\bibitem[{Kuwata et~al.(2024)Kuwata, Yagasaki, and Suga}]{kuwata_2024}
\bibinfo{author}{Y.~Kuwata}, \bibinfo{author}{W.~Yagasaki},
  \bibinfo{author}{K.~Suga},
\newblock \bibinfo{title}{Effects of steepness on turbulent heat transfer over
  sinusoidal rough surfaces},
\newblock \bibinfo{journal}{Int. J. Heat Fluid Flow} \bibinfo{volume}{109}
  (\bibinfo{year}{2024}) \bibinfo{pages}{109537}.
  \DOIprefix\doi{10.1016/j.ijheatfluidflow.2024.109537}.
\bibitem[{Rowin et~al.(2024)Rowin, Zhong, Saurav, Jelly, Hutchins, and
  Chung}]{rowin_2024}
\bibinfo{author}{W.~A. Rowin}, \bibinfo{author}{K.~Zhong},
  \bibinfo{author}{T.~Saurav}, \bibinfo{author}{T.~Jelly},
  \bibinfo{author}{N.~Hutchins}, \bibinfo{author}{D.~Chung},
\newblock \bibinfo{title}{Modelling the effect of roughness density on
  turbulent forced convection},
\newblock \bibinfo{journal}{J. Fluid Mech.} \bibinfo{volume}{979}
  (\bibinfo{year}{2024}) \bibinfo{pages}{A22}.
  \DOIprefix\doi{10.1017/jfm.2023.1063}.
\bibitem[{Peeters and Sandham(2019)}]{peeters_2019}
\bibinfo{author}{J.~Peeters}, \bibinfo{author}{N.~Sandham},
\newblock \bibinfo{title}{Turbulent heat transfer in channels with irregular
  roughness},
\newblock \bibinfo{journal}{Int. J. Heat Mass Transfer} \bibinfo{volume}{138}
  (\bibinfo{year}{2019}) \bibinfo{pages}{454--467}.
  \DOIprefix\doi{10.1016/j.ijheatmasstransfer.2019.04.013}.
\bibitem[{Kuwata(2021)}]{kuwata_2021}
\bibinfo{author}{Y.~Kuwata},
\newblock \bibinfo{title}{Direct numerical simulation of turbulent heat
  transfer on the {Reynolds} analogy over irregular rough surfaces},
\newblock \bibinfo{journal}{Int. J. Heat Fluid Flow} \bibinfo{volume}{92}
  (\bibinfo{year}{2021}) \bibinfo{pages}{108859}.
  \DOIprefix\doi{10.1016/j.ijheatfluidflow.2021.108859}.

\end{thebibliography}
